\documentclass[preprint,aps,prd]{revtex4-1}
\usepackage{graphicx}
\usepackage{amssymb}
\usepackage{slashed}

\newcommand{\be}{\begin{eqnarray}}
\newcommand{\ee}{\end{eqnarray}}
\usepackage[colorlinks]{hyperref}
\usepackage{cancel,bm}

\usepackage{amsmath,amssymb}
\usepackage{mathtools}
\usepackage{float}
\usepackage{color}
\usepackage{leftidx}
\usepackage{booktabs}
\usepackage{latexsym}
\usepackage{revsymb}
\usepackage{multirow}
\usepackage{hypcap}
\usepackage{array}
\hypersetup{
    colorlinks=red,
    citecolor=blue,
    filecolor=magenta,      
    urlcolor=blue,
}

\begin{document}

\title{$J/\psi$ Production in Polarized and Unpolarized $ep$ Collision and  Sivers and $\cos2\phi$ 
Asymmetries }

\author{Asmita Mukherjee and Sangem Rajesh}

    
\affiliation{ Department of Physics,
Indian Institute of Technology Bombay, Mumbai-400076,
India.}
\date{\today}

\begin{abstract}

We calculate the Sivers and $\cos2\phi$ azimuthal asymmetries in $J/\psi$ production in polarized and 
unpolarized  semi-inclusive $ep$ collision respectively, using the
formalism based on transverse momentum dependent parton distributions (TMDs). Non-relativistic 
QCD based color octet model is employed for calculating the $J/\psi$ production rate. The 
Sivers asymmetry in this process directly probes the gluon Sivers 
function. The estimated Sivers asymmetry at $z=1$ is  negative  which is in good 
agreement with COMPASS data.  We also investigate the effect of TMD evolution on the  Sivers asymmetry. The
$\cos2\phi$ asymmetry is sizable and probes the linearly polarized gluon distribution in an unpolarized 
proton.
\end{abstract}

\maketitle
\raggedbottom 
\section{Introduction}\label{sec1}
Single spin asymmetry (SSA) has been playing a vital role in spin physics since the observation of large SSA
in high energy $pp$ collision experimentally \cite{ssa1,ssa2,ssa3,ssa4,ssa5}. 
SSA arises in scattering process in which the target or one of the colliding proton is transversely polarized with
respect to the scattering plane. In order to explain the SSA theoretically, it requires the nonperturbative quark or 
gluon correlators and there are two approaches for it. First one is based on generalized factorization
\cite{jcollins}, where one includes intrinsic transverse momentum in the parton distribution functions and 
fragmentation functions (TMDs). This approach is applicable when the process involves two scales, namely a hard 
and a soft scale. Example of such process is semi inclusive deep inelastic scattering ( SIDIS), where the hard scale is 
the virtuality  of the gauge boson exchanged and the soft scale can be characterized by the transverse momentum of 
the observed hadron. Another such process is Drell-Yan (DY), where the hard scale is the same as SIDIS and the soft 
scale is the transverse momentum of the lepton pair produced. This approach is phenomenologically well studied
\cite{Ji:2004xq,Ji:2004wu,Echevarria:2011rb,Bacchetta:2006tn,Anselmino:2002pd,Boer:1999mm,Arnold:2008kf,Boer:1997mf,Anselmino:2007fs}.
The second approach describes the SSAs in  terms of collinear higher twist quark-gluon correlators.
This formalism uses collinear factorization and was originally proposed in \cite{Efremov:1981sh,Efremov:1984ip,Qiu:1991pp,Qiu:1998ia,Kanazawa:2000hz}
and further developed by \cite{Kouvaris:2006zy,Eguchi:2006mc,Kanazawa:2014dca}. This  is useful for 
processes having only one hard scale like SSA in $pp$ collision.\par
Among the single spin asymmetries, the Sivers asymmetry is one of the most important and well studied asymmetry,
both theoretically and experimentally. This asymmetry involves the Sivers function \cite{Sivers:1989cc}. The 
asymmetry arises because the distribution of quarks and gluons in a transversely polarized proton
is not left-right symmetric with respect to the plane formed by its transverse momentum and spin direction.
The Sivers effect leads to an asymmetry in the azimuthal angle of the hadron produced in SIDIS and has been 
observed in HERMES \cite{Airapetian:2004tw,Airapetian:2009ae} and COMPASS experiments \cite{Adolph:2012sp,Qian:2011py} for proton target 
and by JLab Hall-A collaboration for $\leftidx{^{3}}{\mathrm{He}}{}$ target \cite{Zhao:2014qvx}. The Sivers
function has been shown in a model dependent way to be related to the orbital angular momentum of the quarks 
and gluons \cite{Burkardt:2003uw,Burkardt:2003je}. The first transverse moment of the Sivers function is related
to the quark-gluon twist three Qiu-Sterman function \cite{Boer:2003cm}. A detailed discussion of such relations
 can be found in \cite{Boer:2015vso}. \par
 Sivers function is a T-odd (time reversal odd) object . The operator definitions of the quark and gluon Sivers
 function need gauge links (one for quark Sivers function and two for gluon Sivers function) for color gauge 
 invariance. As these gauge links or Wilson lines depend on the specific process under consideration, this 
 introduces non-universality or process dependence in the Sivers function \cite{Boer:2003cm}. For gluon Sivers
 function, there are two gauge links and the process dependence is more involved. However, the gluon Sivers 
 function for any process can be written in terms of two "universal" gluon Sivers functions \cite{Buffing:2013kca},
 one involving a C-even operator (f-type) , the other a C-odd operator (d-type).\par
 Gluon Sivers function (GSF) plays an important role in 
 understanding the SSAs observed in $pp$ collision as well as those in SIDIS over a wide kinematical region.
 What is more interesting is that different experiments probe different gluon Sivers functions. Burkardt's
 sum rule \cite{Burkardt:2004ur} gives a bound on the GSF. This sum rule is derived from 
the  fact that the total transverse momentum of all partons in a transversely polarized proton should 
vanish.
 Fits to SIDIS data at low scale have found that this sum rule is almost saturated by contribution  from the 
 $u$ and $d$ quark's Sivers function \cite{Anselmino:2008sga}, however there is still room for about  $30\%$ contribution
 from GSF. Moreover, one of the gluon Sivers functions (d-type) is not constrained by the 
 Burkardt's sum rule.
  Apart from SIDIS and DY \cite{Anselmino:2008sga,Anselmino:2009st,Anselmino:2005ea}, Sivers effect has been 
 studied theoretically in several $ep^\uparrow$ collision processes, among them photoproduction of $J/\psi$  \cite{Godbole:2012bx,Godbole:2013bca,Godbole:2014tha},  heavy quark pair and dijet production
 in $ep^{\uparrow}$ scattering \cite{Boer:2016fqd}. In SSA in proton-proton collision, the process dependent initial and final state
 interactions play a major role and usually need to be carefully taken into account \cite{DAlesio:2017rzj}.  \par
 $J/\psi$ production in $ep^{\uparrow}$ scattering provides direct access to the GSF (f-type)  through 
 the leading order (LO) subprocess.  It has been shown
 that \cite{Yuan:2008vn}, due to the final state interactions in $ep$ and $pp$ scattering process, SSA in heavy
 quarkonium production is zero in $ep$ scattering when the heavy quark pair is produced in a color singlet state, 
 whereas for $pp$ scattering the SSA is zero when the heavy quark pair is produced in color octet
 state. Quarkonium production has been studied in unpolarized $pp$ scattering within TMD evolution formalism in \cite{Mukherjee:2015smo,Mukherjee:2016cjw}. In Ref. 
\cite{Godbole:2012bx,Godbole:2013bca,Godbole:2014tha}, SSA in $J/\psi$
 production in $ep^{\uparrow}$ collision using low virtuality electroproduction approximation 
(photoproduction)  is studied in color evaporation model (CEM)  and sizable asymmetries are reported.
 In this work, we investigate the Sivers asymmetry in the semi-inclusive process $e+p^{\uparrow}\to e+J/\psi 
+X$ and the $\cos2\phi$ azimuthal asymmetry in the unpolarized process  $e+p\to e+J/\psi+X$  using  
non-relativistic 
Quantum Chromo Dynamics (NRQCD) based color octet model (COM) \cite{cacciari}.  In COM, the $c\bar{c}$ pair 
is produced in the color octet state that forms $J/\psi$ by emitting soft gluons \cite{Bodwin:1994jh}. The COM 
 is based on a factorization formula in NRQCD. The cross section is described in terms of a product of 
a perturbative part, where the initial state partons form a $c\bar{c}$ pair having definite color and total 
angular momentum quantum numbers, and a non-perturbative matrix element through which the $c\bar{c}$ pair forms
$J/\psi$. These matrix elements are obtained by fitting data and they are universal. We use a recent 
extraction \cite{alesio} for the gluon Sivers function from the SSA data in $pp$ collision at RHIC.\par
The TMDs (unpolarized as well as the Sivers function) depend on the scale, as a result the SSA also depends 
on the scale \cite{Aybat:2011ta}. The scale dependence is given by the TMD evolution and is usually performed in 
the impact parameter or $b_\perp$-space \cite{aybat,Aybat:2011zv}. There are different schemes of performing the TMD evolution, and 
an improved evolution scheme  called CSS2 has been proposed. A detailed discussion of the evolution schemes and scheme
 transformation issues are discussed in the recent paper \cite{Collins:2017oxh}. The evolution in the renormalization scale and
rapidity scales are performed using renormalization group  and Collins-Soper (CS) equations.
To incorporate the correct evolution at large $b_\perp$ value  a nonperturbative Sudakov factor is 
included in the evolution which is usually obtained by fitting the data. We also study the effect of TMD 
evolution on the Sivers asymmetry in $J/\psi$ production in COM. \par
The $\cos2\phi$ azimuthal asymmetry was observed experimentally long ago both in unpolarized 
SIDIS \cite{Arneodo:1986cf,Breitweg:2000qh} and DY \cite{Falciano:1986wk,Guanziroli:1987rp} processes.  
Recently, HERMES \cite{Airapetian:2012yg} and COMPASS \cite{Adolph:2014pwc} experiments reported sizable 
azimuthal asymmetries in low transverse momentum region. In  \cite{Boer:1999mm} it was suggested that the 
$\cos2\phi$ asymmetry could be explained by the Boer-Mulders effect. The   $\cos2\phi$ 
asymmetry arises in the unpolarized cross section due to the correlation between the transverse spin and 
transverse momentum of the parton inside the nucleon. As a result, Boer-Mulders TMD function appears along 
with $\cos2\phi$ term in the unpolarized cross section. Quark (anti-quark) version Boer-Mulders 
function, $h_1^{\perp q}$ (T-odd), represents the transversely polarized quark (anti-quark) distribution 
inside an unpolarized hadron. $h_1^{\perp q}$ has been extracted in 
\cite{Barone:2009hw,Barone:2015ksa,Barone:2006ws} 
from $\cos2\phi$ asymmetry SIDIS data assuming a relation with Sivers function. However, gluon 
Boer-Mulders function, $h_1^{\perp g}$ (T-even), has not been extracted yet.  $h_1^{\perp g}$ represents 
the linearly polarized gluon distribution inside an unpolarized hadron. $\cos2\phi$ asymmetry in the 
production of $J/\psi$ in unpolarized semi-inclusive $ep$ collision process directly allows us 
to probe $h_1^{\perp g}$. The paper is organized as follows. 
Sivers asymmetry and TMD evolution are presented in Sec.\ref{sec2} and Sec.\ref{sec3} 
respectively. Sec.\ref{sec4} and Sec.\ref{sec5} discuss  the $\cos2\phi$ azimuthal asymmetry and 
numerical 
results  respectively along with  the conclusion in Sec.\ref{sec6}.
\section{Sivers asymmetry}\label{sec2}
\begin{figure}[H]
\begin{center} 
\includegraphics[width=12cm,height=10.5cm,clip]{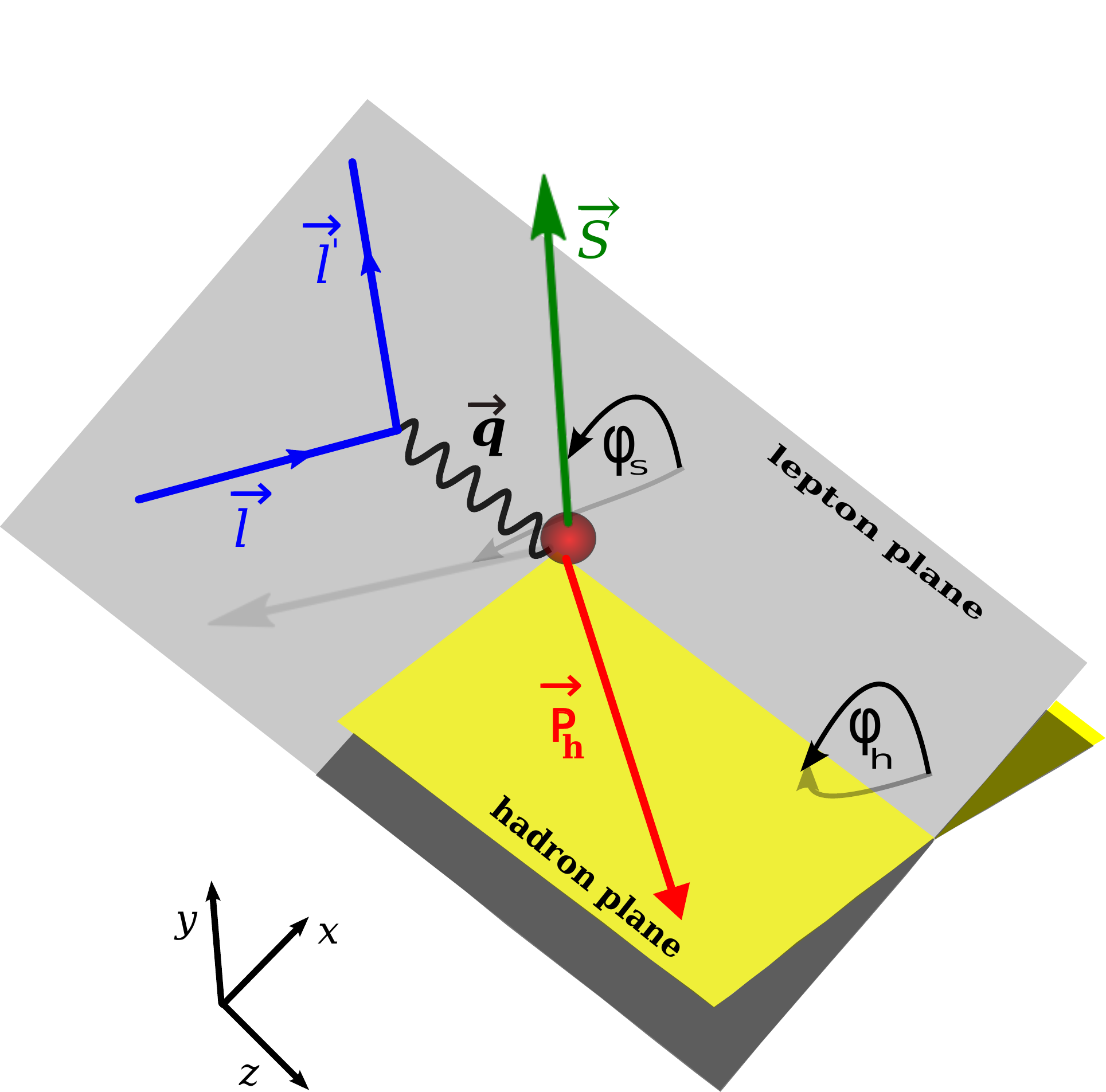}
\end{center}
\caption{\label{fig1}Definition of azimuthal angles ($\phi_s,~\phi_h$), lepton and hadron scattering planes 
in semi inclusive deep inelastic scattering.}
\end{figure}

Single spin asymmetry  for the semi-inclusive process $A^{\uparrow}+ B\to C + X$ is defined as
\begin{eqnarray}\label{asy}
 A_{N}=\frac{d \sigma^{\uparrow}-d \sigma^{\downarrow}}{d \sigma^{\uparrow}+d \sigma^{\downarrow}},
\end{eqnarray}
where $d\sigma^{\uparrow}$ and $d\sigma^{\downarrow}$ are respectively the  differential cross-sections 
measured when one of the particle is transversely polarized up ($\uparrow$) and down ($\downarrow$) with 
respect to the scattering plane. We consider the process, 
\be
e(l)+p^{\uparrow}(P)\rightarrow e(l^\prime)+J/\psi(P_h) +X,
\ee
where the electron scatters by the transversely polarized proton target. The letters within the 
brackets represent the four momentum of the corresponding particle. We follow the generalized 
factorization theorem where the intrinsic partonic transverse 
momentum is taken into account unlike  the collinear factorization. The kinematics considered below are different from 
\cite{Godbole:2012bx,Godbole:2013bca,Godbole:2014tha}.  We consider the frame as shown in 
\figurename{\ref{fig1}}, in which the 
proton and virtual photon are moving along $-z$ and $+z$ axes respectively. The four momenta of target 
system $P$ and virtual photon $q=l-l^\prime$ are given by
\be
P=n_-+\frac{M_p^2}{2}n_+\approx n_-~~\mathrm{and}~~q=-x_Bn_-+\frac{Q^2}{2x_B}n_+\approx -x_BP+(P.q)n_+,
\ee
with $Q^2=-q^2$ and Bjorken variable, $x_B=\frac{Q^2}{2P.q}$ (up to proton mass correction). Here, $M_p$ is 
 mass of the 
proton. The 
leptonic four momenta are expanded in terms of $n_-=P$ and $n_+=n=(q+x_BP)/P.q$  \cite{Pisano:2013cya} as 
follows
\be
l=\frac{1-y}{y}x_BP+\frac{1}{y}\frac{Q^2}{2x_B}n+\frac{\sqrt{1-y}}{y}Q\hat{l}_\perp=\frac{1-y}{y}
x_BP+\frac{s}{2}n+\frac{\sqrt{1-y}}{y}Q\hat{l}_\perp,
\ee
\be
l^\prime=\frac{1}{y}x_BP+\frac{1-y}{y}\frac{Q^2}{2x_B}n+\frac{\sqrt{1-y}}{y}Q\hat{l}_\perp=\frac{1}{y}
x_BP+(1-y)\frac{s}{2}n+\frac{\sqrt{1-y}}{y}Q\hat{l}_\perp,
\ee
here, $y=\frac{P.q}{P.l}$. The invariant mass of electron-target system is 
$s=(P+l)^2=2P.l=\frac{2P.q}{y}$ and then we have $Q^2=x_Bys$. The virtual photon-target invariant mass is 
defined as $W^2=(q+P)^2=\frac{Q^2(1-x_B)}{x_B}$. Using Sudakov decomposition, the four momenta of 
the initial gluon $k$ and the final hadron $P_h$ are 
\be
k=xP+k_\perp+\frac{{\bm k}^2_\perp}{2x}n\approx xP+k_\perp,
\ee
\be
P_h=z(P.q)n+\frac{M^2+{\bm P}_{hT}^2}{2zP.q}P+P_{hT},
\ee
where,  $x=k.n$ is the longitudinal momentum fraction, $z=P.P_h/P.q$ and $P^2_{hT}=-{\bm P}^2_{hT}$. 
Mass of the $J/\psi$ is denoted with $M$. In line with Ref. \cite{Pisano:2013cya} , we assume that 
generalized factorization theorem allows to factorize the unpolarized differential cross section as 
\be\label{dsigma1}
d\sigma=\frac{1}{2s}\frac{d^3l^\prime}{(2\pi)^32E^\prime_l}\frac{d^3P_h}{(2\pi)^32E_h}\int dx d^2{\bm 
k}_\perp (2\pi)^4\delta^4(q+k-P_h)\nonumber\\
\times \frac{1}{Q^4}L^{\nu\nu^\prime}(l,q)\Phi^{\mu\mu^\prime}_{g}(x,{\bm k}_\perp)
\mathcal{M}_{\mu\nu}^{\gamma^\ast g\rightarrow J/\psi}\mathcal{M}_{\mu^\prime\nu^\prime}^{\ast~\gamma^\ast 
g\rightarrow J/\psi}
\ee
The leptonic tensor is given by 
\be\label{lt}
L^{\nu\nu^\prime}(l,q)=e^2\left(-g^{\nu\nu^\prime}Q^2+2(l^\nu l^{\prime\nu^\prime}+l^{\nu^\prime} l^{\prime 
\nu})\right).
\ee
 The gluon-gluon correlator, $\Phi_{g}^{\mu\mu^\prime}(x,{\bm k}_\perp)$, describes the hadron to parton 
transition 
which is parametrized in terms of eight TMDs at leading twist. The gluon correlator is defined 
for unpolarized and transversely polarized hadron respectively  as below \cite{Mulders:2000sh}
\be
\Phi_{g}^{\mu\mu^\prime}(x,{\bm k}_\perp)=\frac{1}{2x}\Bigg\{-g_{T}^{\mu\mu^\prime}f_1^g(x,{\bm 
k}^2_\perp)+\left(\frac{k_{\perp}^{\mu} k_{\perp}^{\mu^\prime}}{M_p^2}+g_{T}^{\mu\mu^\prime}\frac{{\bm 
k}^2_\perp}{2M_p^2}\right)h^{\perp g}_1(x,{\bm k}^2_\perp)\Bigg\},
\ee
\be
\Phi_{g}^{T\mu\mu^\prime}(x,{\bm 
k}_\perp)=-\frac{1}{2x}g_{T}^{\mu\mu^\prime}\frac{\epsilon_T^{\rho\sigma}k_{\perp \rho}S_{T\sigma}}{M_p} 
f_{1T}^{\perp g}(x,{\bm 
k}^2_\perp)
\ee
where $g_{T}^{\mu\mu^\prime}=g^{\mu\mu^\prime}-P^{\mu}n^{\mu^\prime}/P.n-P^{\mu^\prime}n^{\mu}/P.n$ is the 
transverse metric tensor.  Here we have kept only the part of the hadronic tensor for transverse polarization, that contributes to the Sivers asymmetry. $f_1^g$ and $h_1^{\perp g}$ represent the unpolarized and 
linearly polarized gluon distribution functions inside the unpolarized hadron respectively. $f_{1T}^{\perp 
g}$, gluon Sivers function, describes the density of unpolarized gluons inside the transversely polarized 
hadron.
The only LO subprocess for $J/\psi$ production is $\gamma^\ast g\rightarrow c\bar{c}$.
In Eq.\eqref{dsigma1}, $\mathcal{M}^{\gamma^\ast g\rightarrow J/\psi}$ is the amplitude of $J/\psi$ 
production.  $J/\psi$ production mechanism, for instance, contains both perturbative  and 
nonperturbative regimes which need to be separated out systematically. We employ the COM to calculate 
the amplitude of $J/\psi$ bound state.
The detailed calculation is discussed in the Appendix.  In COM framework, initially 
heavy quark pair  produced in a definite quantum state which can be calculated using perturbation theory 
up to a fixed order in $\alpha_s$.  The long distance matrix element (LDME), $\langle 0\mid 
\mathcal{O}^{J/\psi}_n\mid 0\rangle$, contains the transition probability of  $J/\psi$ production from heavy 
quark pair.   The momentum conservation delta function can be decomposed as 
\be\label{delta}
\delta^4(q+k-P_h)=\frac{2}{ys}\delta\left(x-x_B-\frac{M^2+{\bm P}_{hT}^2}{zys}\right)\delta(1-z)\delta^2({\bm 
k}_\perp-{\bm P}_{hT}).
\ee
The phase space factors in Eq.\eqref{dsigma1} can be written as follows
\be\label{ps}
\frac{d^3l^\prime}{(2\pi)^32E^\prime_l}=\frac{1}{16\pi^2}sydx_Bdy,~~~
\frac{d^3P_h}{(2\pi)^32E_h}=\frac{dzd^2{\bm P}_{hT}}{(2\pi)^32z}.
\ee
The differential cross section can be expressed in terms of TMDs by substituting parameterization of 
gluon correlator, the leptonic tensor and Eq.\eqref{a13}-\eqref{a17}  in Eq.\eqref{dsigma1}. Using 
Eq.\eqref{lt}-\eqref{ps} and after integrating with respect 
to $x$ and $z$, one obtains
\be\label{unp1}
\frac{d\sigma}{dydx_Bd^2{\bm 
P}_{hT}}=\frac{\alpha}{8sxQ^4}\int d^2{\bm k}_\perp \left[\mathcal{A}_0+\mathcal{A}_1\cos 
\phi\right]f_{g/p}(x,{\bm 
k}_{\perp}^2)\delta^2({\bm k}_\perp-{\bm P}_{hT}),
\ee
with correction  $\mathcal{O}\left(\frac{k_\perp^2}{(M^2+Q^2)^2}\right)$. The azimuthal angle of the initial 
gluon transverse momentum is denoted with $\phi$. For obtaining Eq.\eqref{unp1}, $\phi=\phi_h$ is understood 
where $\phi_h$ is the azimuthal angle of the $J/\psi$.  In Eq.\eqref{unp1}, only the unpolarized gluon 
contribution is 
taken into 
consideration. The effect of linearly 
polarized gluon contribution will be discussed in the Sec.\ref{sec4}. 
We define $\mathcal{A}_0$ and $\mathcal{A}_1$ as
\begin{eqnarray}\label{da0}
\begin{aligned}
\mathcal{A}_0={}&[1+(1-y)^2]\frac{\mathcal{N}Q^2}{y^2M}\Bigg\{\langle 0\mid 
\mathcal{O}_8^{J/\psi}(\leftidx{^1}{S}{_0})\mid 0\rangle+\frac{4}{3M^2}\frac{ (3 M^2+Q^2)^2}{(M^2+Q^2)^2}
\langle 0\mid \mathcal{O}_8^{J/\psi}(\leftidx{^3}{P}{_0})\mid 0\rangle\\
&+\frac{8 Q^2}{3M^2(M^2+Q^2)^2} 
 \left(\frac{4M^2 (1-y)}{1+(1-y)^2}+Q^2\right)\langle 0\mid 
\mathcal{O}_8^{J/\psi}(\leftidx{^3}{P}{_1})\mid 0\rangle\\
&+\frac{8 }{15M^2  \left(M^2+Q^2\right)^2}\left(6 M^4+Q^4+12 M^2 Q^2 \frac{1-y}{1+(1-y)^2}\right)
\langle 0\mid \mathcal{O}_8^{J/\psi}(\leftidx{^3}{P}{_2})\mid 0\rangle\Bigg\}
\end{aligned}
\end{eqnarray}
\begin{eqnarray}\label{da1}
\begin{aligned}
\mathcal{A}_1={}&(2-y)\sqrt{1-y}\frac{4\mathcal{N}Q^3}{y^2M}\Bigg\{- 
\langle 0\mid\mathcal{O}_8^{J/\psi}(\leftidx{^1}{S}{_0})\mid 0\rangle-\frac{2}{3M^2Q^2}\frac{ 
(3 M^2+Q^2)^2}{M^2+Q^2}
\langle 0\mid \mathcal{O}_8^{J/\psi}(\leftidx{^3}{P}{_0})\mid 0\rangle\\
&-\frac{8 Q^2}{3M^2(M^2+Q^2)}\langle 0\mid 
\mathcal{O}_8^{J/\psi}(\leftidx{^3}{P}{_1})\mid 0\rangle
-\frac{4 }{15 M^2}\frac{7M^2+Q^2}{M^2+Q^2}
\langle 0\mid \mathcal{O}_8^{J/\psi}(\leftidx{^3}{P}{_2})\mid 0\rangle
\Bigg\}\frac{k_\perp}{M^2+Q^2},
\end{aligned}
\end{eqnarray}
with $\mathcal{N}=2(4\pi)^2\alpha_s\alpha e_c^2$. $\mathcal{A}_1$ does not contribute to the Sivers asymmetry.
The numerical values of the different states LDME are taken from Ref. \cite{Mukherjee:2016cjw}, Set-I in 
Table-I.
Following Ref. \cite{Anselmino:2005nn},  the numerator  term of the Sivers asymmetry is 
given below  when the target proton is transversely polarized   

\begin{equation}\label{numer}
 \begin{aligned}
 \frac{d\sigma^{\uparrow}}{dydx_Bd^2{\bm 
P}_{hT}}-\frac{d\sigma^{\downarrow}}{dydx_Bd^2{\bm P}_{hT}}
 ={}&\frac{\alpha}{8sxQ^4}\left[\mathcal{A}_0+\mathcal{A}_1\cos \phi_h\right] 
\Delta^Nf_{g/p^{\uparrow}}(x,{\bm P}_{hT}).
 \end{aligned}
\end{equation}
The gluon Sivers function 
as per Trento convention is given by \cite{Bacchetta:2004jz}
\be\label{s1}
\Delta^Nf_{g/p^{\uparrow}}(x,{\bm P}_{hT},Q_f)&=&-2f_{1T}^{\perp g}(x,{\bm P}_{hT},Q_f)
\frac{(\hat{\bm P}\times {\bm P}_{hT}).{\bm S}}{M_p}.
\ee
The scale dependency in the definition of TMD  is suppressed in this section. The denominator term is given by
\begin{equation}\label{denom}
 \begin{aligned}
 \frac{d\sigma^{\uparrow}}{dydx_Bd^2{\bm 
P}_{hT}}+\frac{d\sigma^{\downarrow}}{dydx_Bd^2{\bm P}_{hT}}
 ={}&\frac{2\alpha}{8sxQ^4}\left[\mathcal{A}_0+\mathcal{A}_1\cos \phi_h\right] f_{g/p}(x,{\bm P}_{hT}^2),
 \end{aligned}
\end{equation}
where the GSF $\Delta^Nf$ describes the probability of finding an unpolarized gluon 
inside a transversely polarized proton which is defined as
\be
\Delta^Nf_{g/p^{\uparrow}}(x,{\bm P}_{hT})
&=& f_{g/p^{\uparrow}}(x,{\bm P}_{hT})- f_{g/p^{\downarrow}}(x,{\bm P}_{hT})\nonumber\\
&=&\Delta^Nf_{g/p^{\uparrow}}(x,P_{hT})~{\bm S}.(\hat{\bm P}\times\hat{\bm 
P}_{hT}).
\ee

\section{Evolution of TMDs}\label{sec3}
In this section the evolution of TMDs is studied.
 It is generally assumed that the unpolarized  gluon TMDs obey the Gaussian distribution. The 
Gaussian parameterization of unpolarized TMD is given by
\be \label{unp}
 f_{g/p}(x,{\bm k}^2_{\perp })=f_{g/p}(x,\mu)\frac{1}{\pi \langle k^2_{\perp }\rangle}
 e^{-{\bm k}^2_{\perp }/\langle k^2_{\perp }\rangle}.
\ee
Here, $x$ and $k_\perp$ dependencies of the TMD are factorized. $f_{g/p}(x,\mu)$ is the collinear PDF  
which is measured at the scale $\mu=M$ (mass of $J/\psi$). The collinear PDF obeys the 
Dokshitzer-Gribov-Lipatov-Altarelli-Parisi (DGLAP) scale evolution.
We choose a frame where the polarized proton is moving along $-z$ axis with momentum ${\bm P}$ and is 
transversely polarized with $S=S_T(\cos\phi_s,\sin\phi_s,0)$. The transverse momentum of the $J/\psi$
is  ${\bm P}_{hT}=P_{hT}(\cos\phi_h,\sin\phi_h,0)$
\be
 {\bm S}.(\hat{\bm P}\times\hat{\bm P}_{hT})=\sin(\phi_h-\phi_s),
\ee
where, $\phi_s$ and $\phi_h$ are the azimuthal angles which are defined in \figurename{ \ref{fig1}}.
 The parameterization of GSF is given by \cite{alesio,Anselmino:2016uie}
\be
\Delta^{N}f_{g/p^{\uparrow}}(x, k_{\perp})=2\mathcal{N}_g(x)f_{g/p}(x,\mu)h(k_\perp)
\frac{e^{-k^2_\perp/\langle k^2_\perp\rangle}}{\pi\langle k^2_\perp\rangle} 
\ee
here 
\be\label{ngx}
\mathcal{N}_g(x)=N_gx^\alpha(1-x)^\beta\frac{(\alpha+\beta)^{(\alpha+\beta)}}{
\alpha^\alpha\beta^\beta}.
\ee
$h(k_\perp)$ is defined as follows
\be
h(k_\perp)=\sqrt{2e}\frac{k_\perp}{M_1}e^{-k^2_\perp/M^2_1}
\ee
Therefore, the $k_\perp$ dependent part of Sivers function can now be written as
\be
h(k_\perp)\frac{e^{-k^2_\perp/\langle k^2_\perp\rangle}}{\pi\langle k^2_\perp\rangle}
=\frac{\sqrt{2e}}{\pi}\sqrt{\frac{1-\rho}{\rho}}k_\perp
\frac{e^{-k^2_{\perp}/\rho\langle k^2_\perp\rangle}}{\langle k^2_\perp\rangle^{3/2}},
\ee
where we defined 
\be
\rho=\frac{M^2_1}{\langle k^2_\perp\rangle + M^2_1}
\ee
The GSF has been extracted first time in pion production at RHIC \cite{rhic1} by D'Alesio et al. 
\cite{alesio}. In this analysis \cite{alesio}, the best fit parameter sets  are  denoted with SIDIS1 and 
SIDIS2. Recently, M. Anselmino et al. \cite{Anselmino:2016uie} have extracted the quark and 
anti-quark Sivers function from latest SIDIS data. However, GSF has not been extracted yet from SIDIS 
data. Therefore, in order to estimate the asymmetry,  best fit 
parameters of Sivers function corresponding to $u$ and $d$ quark  will be used in 
the following  parameterizations \cite{Boer:2003tx} : 
\be \label{ab}
(a)~~\mathcal{N}_g(x)&=&({N}_u(x)+{N}_d(x))/2 \nonumber\\
(b)~~\mathcal{N}_g(x)&=&{N}_d(x)
\ee
We call the  parameterization (a) and (b) as BV-a and BV-b respectively. The best fit parameters are 
tabulated in \tablename{ \ref{table1}}.
\begin{table}[H]
\centering
\begin{tabular}{ | >{\centering\arraybackslash}m{2cm}| >{\centering\arraybackslash}m{1.2cm}| 
>{\centering\arraybackslash}m{1.2cm}| >{\centering\arraybackslash}m{1.2cm}| 
>{\centering\arraybackslash}m{1.2cm}| >{\centering\arraybackslash}m{1.2cm}| 
>{\centering\arraybackslash}m{2cm}| >{\centering\arraybackslash}m{2cm}| 
>{\centering\arraybackslash}m{1.5cm}| }
\hline
\multicolumn{9}{ |c| }{Best fit parameters} \\
\cline{1-9}
Evolution & $a$ & $N_a$ & $\alpha$ & $\beta$ &$\rho$ &$M_1^2$ GeV$^2$& $\langle k^2_\perp\rangle$ 
GeV$^2$ & 
Notation  \\ \cline{1-9}
\multicolumn{1}{ |c  } {\multirow{4}{*}{DGLAP}} &
\multicolumn{1}{ |c| } {$g$ \cite{alesio}} & 0.65 & 2.8 & 2.8 & 0.687 & & 0.25 & SIDIS1 \\
  \cline{2-9}
  \multicolumn{1}{ |c  }{} &
  \multicolumn{1}{ |c| } {$g$ \cite{alesio}} & 0.05 & 0.8 & 1.4 & 0.576 & & 0.25 & SIDIS2 \\
  \cline{2-9}
  \multicolumn{1}{ |c  }{} &
  \multicolumn{1}{ |c| } {$u$ \cite{Anselmino:2016uie}} & 0.18 & 1.0 & 6.6 &  & 0.8 & 0.57 & BV-a \\
  \cline{2-8}
  \multicolumn{1}{ |c  }{} &
  \multicolumn{1}{ |c| } {$d$ \cite{Anselmino:2016uie}} & -0.52 & 1.9 & 10.0 &  & 0.8 & 0.57 & BV-b \\
  \cline{1-9}
  \multicolumn{1}{ |c  } {\multirow{2}{*}{TMD}} &
  \multicolumn{1}{ |c| } {$u$ \cite{echevarria}} & 0.106 & 1.051 & 4.857 &  &  & 0.38 & TMD-a \\
  \cline{2-8}
  \multicolumn{1}{ |c  }{} &
  \multicolumn{1}{ |c| } {$d$ \cite{echevarria}} & -0.163 & 1.552 & 4.857 &  &  & 0.38 & TMD-b \\
  \cline{1-9}
\end{tabular}
\caption{\label{table1}Best fit parameters of Sivers function.}
\end{table}
We use the nonuniversality property of Sivers function for only SIDIS1 and SIDIS2 parameters since these 
parameters are extracted in DY process \cite{echevarria}
\be
\Delta^{N}_{\mathrm{DY}}f_{g/p^{\uparrow}}(x, k_{\perp})=-
\Delta^{N}_{\mathrm{SIDIS}}f_{g/p^{\uparrow}}(x, k_{\perp})
\ee
Finally, we are in position to write the final expressions of Eq.\eqref{asy} within DGLAP evolution 
formalism.
Using Eq.\eqref{unp}-\eqref{ab}, the $\sin(\phi_h-\phi_s)$  weighted numerator part of Eq.\eqref{asy} is 
given by 

\begin{equation}\label{numer1}
 \begin{aligned}
 \frac{d\sigma^{\uparrow}}{dydx_Bd^2{\bm 
P}_{hT}}-\frac{d\sigma^{\downarrow}}{dydx_Bd^2{\bm P}_{hT}}
 ={}&\frac{\alpha}{8sxQ^4}\left[\mathcal{A}_0+\mathcal{A}_1\cos 
\phi_h\right]2\mathcal{N}_g(x)\frac{\sqrt{2e}}{\pi}
 \sqrt{\frac{1-\rho}{\rho}}P_{hT}\\
& \times \frac{e^{-P_{hT}^2/\rho\langle P^2_{hT}\rangle}}{\langle P^2_{hT}\rangle^{3/2}}
 f_{g/p}(x)\sin^2(\phi_h-\phi_s),
 \end{aligned}
\end{equation}
and the denominator term as follows
\begin{equation}\label{denom1}
 \begin{aligned}
 \frac{d\sigma^{\uparrow}}{dydx_Bd^2{\bm 
P}_{hT}}+\frac{d\sigma^{\downarrow}}{dydx_Bd^2{\bm P}_{hT}}
 ={}&\frac{2\alpha}{8sxQ^4}\left[\mathcal{A}_0+\mathcal{A}_1\cos \phi_h\right] \frac{e^{-P_{hT}^2/\langle 
P^2_{hT}\rangle}}{\pi\langle P^2_{hT}\rangle}
 f_{g/p}(x).
 \end{aligned}
\end{equation}
\par
Now, we adopt the framework implemented in Ref. \cite{echevarria} to study the TMD evolution. In general, 
TMDs 
 are
defined in impact parameter ($b_\perp$)-space as below
\be\label{ub1}
f(x,b_\perp,\mu)=\int d^2{\bm k}_\perp 
e^{-i{\bm b}_\perp.{\bm k}_\perp}f(x,k_\perp,\mu)
\ee
and the inverse Fourier transformation is 
\be\label{ub2}
f(x,k_\perp,\mu)=\frac{1}{(2\pi)^2}\int d^2{\bm b}_\perp 
e^{i{\bm b}_\perp.{\bm k}_\perp}f(x,b_\perp,\mu).
\ee
Generally, TMDs depend on both renormalization scale ($\mu$) and auxiliary  scale ($\xi$) which is 
introduced 
to regularize the light-cone divergences in TMD factorization formalism \cite{jcollins,aybat}. Taking the 
scale evolution with respect to $\mu$
 and $\xi$ the renormalization group (RG) and Collins-Soper (CS) equations are obtained. By solving these 
equations one obtains
 the TMD PDF expression which is evolved from the initial scale $Q_i=c/b_{\ast}(b_{\perp})$ to final scale
 $Q_f=\zeta=M$ \cite{echevarria,jcollins,aybat,Echevarria:2015uaa}
\begin{eqnarray}{\label{pert}}
  f(x,b_{\perp},Q_f,\zeta)=f(x,b_\perp,Q_i)R_{pert}\left(Q_f,Q_i,b_{\ast}\right)
  R_{NP}\left(Q_f,Q_i,b_{\perp}\right).
 \end{eqnarray}
Here, $R_{pert}$ is the perturbative part. The nonperturbative part of 
the TMDs is denoted with $R_{NP}$. The initial scale of the TMDs is 
$Q_i=c/b_{\ast}(b_{\perp})$,
where $c=2e^{-\gamma_\epsilon}$ with $\gamma_\epsilon\approx0.577$. The widely used $b_{\ast}$ 
prescription
 is adopted to avoid hitting the Landau pole by freezing the scale $b_{\perp}$. Here, 
 $b_{\ast}(b_{\perp})=\frac{b_{\perp}}{\sqrt{1+\left(\frac{b_{\perp}}{b_{\mathrm{max}}}\right)^2}}
 \approx b_{\mathrm{max}}$ when $b_{\perp}\rightarrow \infty$ and $b_{\ast}(b_{\perp})\approx
 b_{\perp}$ when $b_{\perp}\rightarrow 0$.
The perturbative evolution kernel is given by
 \be\label{sudakov}
 R_{pert}\left(Q_f,Q_i,b_{\ast}\right)=\mathrm{exp}\Big\{{-\int_{c/b_{\ast}}^{Q_f}\frac{d\mu}
 {\mu}\left(A\log\left(\frac{Q_f^2} {\mu^2}\right)+B\right)}\Big\},
 \ee
where the anomalous dimensions  are denoted with $A$ an $B$ respectively and 
 these have perturbative  expansion that can be written as :
   $$A=\sum_{n=1}^{\infty}\left(\frac{\alpha_s(\mu)}{\pi}\right)^nA_n$$
 and $$B=\sum_{n=1}^{\infty}\left(\frac{\alpha_s(\mu)}{\pi}\right)^nB_n.$$ 
Here  the anomalous dimension coefficients  $A_1=C_A$, 
$A_2=\frac12C_F\left(C_A\left(\frac{67}{18}-\frac{\pi^2}{6}\right)-\frac{5}{9}C_AN_f\right)$ 
 and $B_1=-\frac{1}{2}(\frac{11}{3}C_A-\frac{2}{3}N_f)$.
These coefficients  are derived up to 3-loop level in Ref. \cite{tmde5}. The nonperturbative part is given 
by 
\be
R_{NP}=\mathrm{exp}\Bigg\{-\Big[g_1^{\mathrm{TMD}}+\frac{g_2}{2}
 \log\frac{Q_f}{Q_0}
 \Big]b_{\perp}^2\Bigg\}
\ee
It is known that  \cite{aybat} the derivative of Sivers function, $f^{\prime \perp}
(x,b_{\perp},Q_f)$, follow the same evolution as that of the unpolarized TMD. The TMD evolution equation 
of unpolarized
gluon TMD PDF is

\be\label{eun1}
f_{g/p}(x,b_\perp,Q_f)=f_1^g(x,b_\perp,Q_i)
\mathrm{exp}\Bigg\{{-\int_{c/b_{\ast}}^{Q_f}
\frac{d\mu}{\mu}\left(A\log\left(\frac{Q_f^2}
 {\mu^2}\right)+B\right)}\Bigg\}\nonumber\\
 \times\mathrm{exp}\Bigg\{-\Big[g_1^{\mathrm{pdf}}+\frac{g_2}{2}        
 \log\frac{Q_f}{Q_0}
 \Big]b_{\perp}^2\Bigg\}
\ee
and derivative of gluon Sivers function is
\be\label{esiv1}
f_{1T}^{\prime \perp g }(x,b_\perp,Q_f)=f_{1T}^{\prime \perp g }(x,b_\perp,Q_i)
\mathrm{exp}\Bigg\{{-\int_{c/b_{\ast}}^{Q_f}
\frac{d\mu}{\mu}\left(A\log\left(\frac{Q_f^2}
 {\mu^2}\right)+B\right)}\Bigg\}\nonumber\\
 \times\mathrm{exp}\Bigg\{-\Big[g_1^{\mathrm{sivers}}+\frac{g_2}{2}
 \log\frac{Q_f}{Q_0}
 \Big]b_{\perp}^2\Bigg\}
\ee

The TMD density function at the initial scale, $f_1^{g}(x,b_\perp,Q_i)$,  can be  written as the 
convolution 
 of coefficient function times the regular  collinear PDF \cite{aybat} 
  \be
  f_1^{g}(x,b_\perp,Q_i)=\sum_{i=g,q}\int_x^1\frac{d\hat{x}}{\hat{x}}
  C_{i/g}(x/\hat{x},b_{\perp},\alpha_s,Q_i)
  f_{i/p}(\hat{x},c/b_{\ast})+\mathcal{O}(b_{\perp}\varLambda_{QCD}),
  \ee
 where $C_{i/g}$ is the perturbatively calculated coefficient function which is process independent. 
$C_{i/g}$ is different for each type of TMD PDF.
 The collinear PDF is probed at the scale $c/b_{\ast}$ rather than the scale $\mu$ 
 in contrast to the  DGLAP evolution. The unpolarized and Sivers function TMDs in terms of collinear 
 PDF at leading order in $\alpha_{s}$ are given by \cite{aybat,echevarria}
 \be\label{et5}
  f_1^g(x,b_\perp,Q_i)=f_{g/p}(x,c/b_{\ast})+\mathcal{O}(\alpha_s),
 \ee
 \be\label{et6}
 f_{1T}^{\prime \perp g }(x,b_\perp,Q_i)\simeq\frac{M_pb_\perp}{2}T_{g,F}(x,x,Q_i)
 \ee
 where  $T_{g,F}(x,x,Q_i)$ is the Qiu-Sterman function proportional to collinear
 PDF \cite{Kouvaris:2006zy}
 \be
 T_{g,F}(x,x,Q_i)=\mathcal{N}_g(x)f_{g/p}(x,Q_i)
 \ee
where $\mathcal{N}_g(x)$ definition is given in Eq.\eqref{ngx}. The numerical values of the free 
parameters  are estimated \cite{echevarria} by global fit of SSA in SIDIS process from pion, kaons and 
charged hadrons production at Jlab, HERMES and COMPASS, which are tabulated in \tablename{ \ref{table1}}.
However, only the $u$ and $d$ quark's free parameters are extracted and gluon parameters are not known 
yet. To estimate SSA we use two parameterizations as given in Eq.\eqref{ab}. We call the  
parameterization (a) and (b) as TMD-a and TMD-b respectively.
 The numerical values of best fit parameters are estimated \cite{echevarria} at 
$Q_0=\sqrt{2.4}~\mathrm{GeV}$, $b_{\mathrm{max}}=1.5\mathrm{~GeV^{-1}}$, $g_2=0.16\mathrm{~GeV^2}$  and 
$\langle k^2_{s\perp }\rangle=0.282\mathrm{~GeV^2}$ with $g_1^{\mathrm{pdf}}=\langle k^2_{\perp 
}\rangle/4$ and $g_1^{\mathrm{sivers}}=\langle k^2_{s \perp  } \rangle/4$. 
The gluon Sivers function $f_{1T}^{\perp g}(x,{\bm P}_{hT},Q_f)$ and it's derivative are related by 
Fourier transformation as below \cite{aybat} 
\be\label{ub5}
f_{1T}^{\perp g}(x,{\bm P}_{hT},Q_f)=-\frac{1}{2\pi P_{hT}}\int_0^\infty db_\perp b_\perp 
J_1(P_{hT}b_\perp)f_{1T}^{\prime \perp g}(x,b_\perp,Q_f)
\ee
and the unpolarized gluon TMD is given by
\be\label{ub5}
f_{g/p}(x, P_{hT},Q_f)=\frac{1}{2\pi}\int_0^\infty db_\perp b_\perp 
J_0({P}_{hT}b_\perp)f_{g/p}(x,b_\perp,Q_f)
\ee
Using above expressions, the  Eq.\eqref{numer},   including the weight factor  $\sin(\phi_h-\phi_s)$ and \eqref{denom}   in TMD evolution framework
can be written as  follows
\begin{equation}\label{numer2}
 \begin{aligned}
 \frac{d\sigma^{\uparrow}}{dydx_Bd^2{\bm 
P}_{hT}}-\frac{d\sigma^{\downarrow}}{dydx_Bd^2{\bm P}_{hT}}
 ={}&\frac{\alpha}{8\pi sxQ^4M_p}
\int_0^\infty db_\perp b_\perp  J_1(P_{hT}b_\perp)
f_{1T}^{\prime \perp g}(x,b_\perp,Q_f)\\
&\times \sin^2(\phi_h-\phi_s)\left[\mathcal{A}_0+\mathcal{A}_1\cos 
\phi_h\right],
 \end{aligned}
\end{equation}
\begin{equation}\label{denom2}
 \begin{aligned}
 \frac{d\sigma^{\uparrow}}{dydx_Bd^2{\bm 
P}_{hT}}+\frac{d\sigma^{\downarrow}}{dydx_Bd^2{\bm P}_{hT}}
 ={}&\frac{\alpha}{8\pi sxQ^4}
 \int_0^\infty db_\perp b_\perp  J_0(P_{hT}b_\perp)
f_{g/p}(x,b_\perp,Q_f)\\
&\times\left[\mathcal{A}_0+\mathcal{A}_1\cos 
\phi_h\right],
 \end{aligned}
\end{equation}
\section{$\cos2\phi$ azimuthal asymmetry}\label{sec4}
Now, let's consider the unpolarized process i.e., $e(l)+p(P)\rightarrow 
e(l^\prime)+J/\psi(P_h) +X$. Taking into account the linearly polarized gluons along with the unpolarized 
gluons in the gluon correlator, the Eq.\eqref{unp1} can be 
written as
\begin{equation}\label{c1}
 \begin{aligned}
\frac{d\sigma}{dydx_Bd^2{\bm 
P}_{hT}}={}&\frac{\alpha}{8sxQ^4}\int d^2{\bm k}_\perp\Big\{ \left[\mathcal{A}_0+\mathcal{A}_1\cos 
\phi\right]f_{g/p}(x,{\bm k}_{\perp}^2)\\
&+{\bm k}_\perp^2\left[\mathcal{B}_0\cos2\phi+\mathcal{B}_1\cos\phi\right] h_1^{\perp g}(x,{\bm 
k}^2_\perp)\Big\}\delta^2({\bm k}_\perp-{\bm P}_{hT}),
\end{aligned}
\end{equation}
with correction $\mathcal{O}\left(\frac{k_\perp^2}{(M^2+Q^2)^2}\right)$. The definitions of $\mathcal{A}_0$ 
and $\mathcal{A}_1$ are given in Eq.\eqref{da0} and Eq.\eqref{da1} respectively.  The $\mathcal{B}_0$ and 
$\mathcal{B}_1$ are defined as below
\begin{eqnarray}\label{bb0}
\begin{aligned}
\mathcal{B}_0={}&(1-y)\frac{\mathcal{N}Q^2}{y^2M}\Bigg\{-\langle 
0\mid\mathcal{O}_8^{J/\psi}(\leftidx{^1}{S}{_0})\mid 0\rangle
+\frac{4}{3M^2}\frac{  \left(3 M^2+Q^2\right)^2 }{ \left(M^2+Q^2\right)^2}
\langle 0\mid \mathcal{O}_8^{J/\psi}(\leftidx{^3}{P}{_0})\mid 0\rangle\\
&-\frac{8 Q^4}{3M^2(M^2+Q^2)^2} 
\langle 0\mid \mathcal{O}_8^{J/\psi}(\leftidx{^3}{P}{_1})\mid 0\rangle 
+\frac{8 Q^4}{15 M^2 \left(M^2+Q^2\right)^2} 
\langle 0\mid \mathcal{O}_8^{J/\psi}(\leftidx{^3}{P}{_2})\mid 0\rangle
\Bigg\}
\end{aligned}
\end{eqnarray}

\begin{eqnarray}\label{bb1}
\begin{aligned}
\mathcal{B}_1={}&(2-y)\sqrt{1-y}\frac{2\mathcal{N}Q}{y^2M} \Bigg\{Q^2\langle 
0\mid\mathcal{O}_8^{J/\psi}(\leftidx{^1}{S}{_0})\mid 0\rangle-
\frac{2}{3M^2}\frac{  \left(3 M^2+Q^2\right)^2 }{M^2+Q^2}
\langle 0\mid \mathcal{O}_8^{J/\psi}(\leftidx{^3}{P}{_0})\mid 0\rangle\\
&+\frac{8 Q^4}{3M^2(M^2+Q^2)}\langle 0\mid \mathcal{O}_8^{J/\psi}(\leftidx{^3}{P}{_1})\mid 0\rangle
-\frac{4Q^2}{15M^2}\frac{Q^2-5M^2}{ M^2+Q^2} 
\langle 0\mid \mathcal{O}_8^{J/\psi}(\leftidx{^3}{P}{_2})\mid 0\rangle
\Bigg\}\frac{k_\perp}{M^2+Q^2}
\end{aligned}
\end{eqnarray}
The dependence of the cross section on azimuthal angle vanishes when  intrinsic parton transverse 
momentum $k_\perp=0$. The $\cos2\phi$ asymmetry is defined as 
\cite{Airapetian:2012yg, Adolph:2014pwc}
\be\label{c2}
<\cos2\phi>=\frac{\int d\phi_h \cos(2\phi_h)d\sigma}{\int d\phi_h d\sigma}.
\ee
To estimate the $\cos2\phi$ asymmetry, we need the parameterization of TMDs. For unpolarized TMD, we 
follow the Gaussian parameterization as defined in Eq.\eqref{unp}. The widely used Gaussian 
parameterization for linearly polarized gluon distribution function is given by \cite{Boer:2012bt}
\be\label{hg}
h_1^{\perp g}(x,{\bf k}^2_{\perp })=\frac{M_p^2f_1^g(x,Q^2)}{\pi\langle k^2_{\perp 
}\rangle^2}\frac{2(1-r)}{r}e^{1-
 {\bf k}^2_{\perp }\frac{1}{r\langle k^2_{\perp }\rangle}},
\ee
where, $r$ ($0<r<1$) is the parameter.  The upper bound on $h_1^{\perp g}$  is given by 
\cite{Boer:2010zf}
\be
\frac{{\bf k}_{\perp}^2}{2M_p^2}|h^{\perp }_1(x,{\bf k}_{\perp }^2)|\leq f_1^g(x,{\bf k}^2_{\perp }).
\ee
We consider $\langle k^2_{\perp }\rangle=0.25$ GeV$^2$  \cite{Boer:2012bt} and  
$r=\frac13$  and  $\frac23$  \cite{Boer:2012bt} for numerical estimation.

\section{Numerical Results}\label{sec5} 

We have estimated the Sivers and $\cos2\phi$ asymmetries respectively  in polarized and unpolarized 
SIDIS processes   using TMD factorization formalism at $\sqrt{s}=4.7\mathrm{~GeV}$ (JLab),  
$\sqrt{s}=7.2\mathrm{~GeV}$ (HERMES),  $\sqrt{s}=17.33\mathrm{~GeV}$ (COMPASS) and  
$\sqrt{s}=45.0\mathrm{~GeV}$ (EIC). In this work,  NRQCD color octet model (COM) is used 
 for $J/\psi$ production. The  color octet states $\leftidx{^{1}}{S}{_0}$, 
 $\leftidx{^{3}}{P}{_0}$, $\leftidx{^{3}}{P}{_1}$  
 and $\leftidx{^{3}}{P}{_2}$ are taken into account for the LO subprocess 
$\gamma^\ast g\rightarrow c\bar{c}$ of charmonium production. $M=3.096\mathrm{~GeV}$ and 
$m_c=1.4\mathrm{~GeV}$ are considered for $J/\psi$ and charm quark mass respectively. MSTW2008 
\cite{mstw} is used for collinear PDFs. \par
\begin{figure}[H]
\begin{minipage}[c]{0.99\textwidth}
\small{(a)}\includegraphics[width=8cm,height=6.5cm,clip]{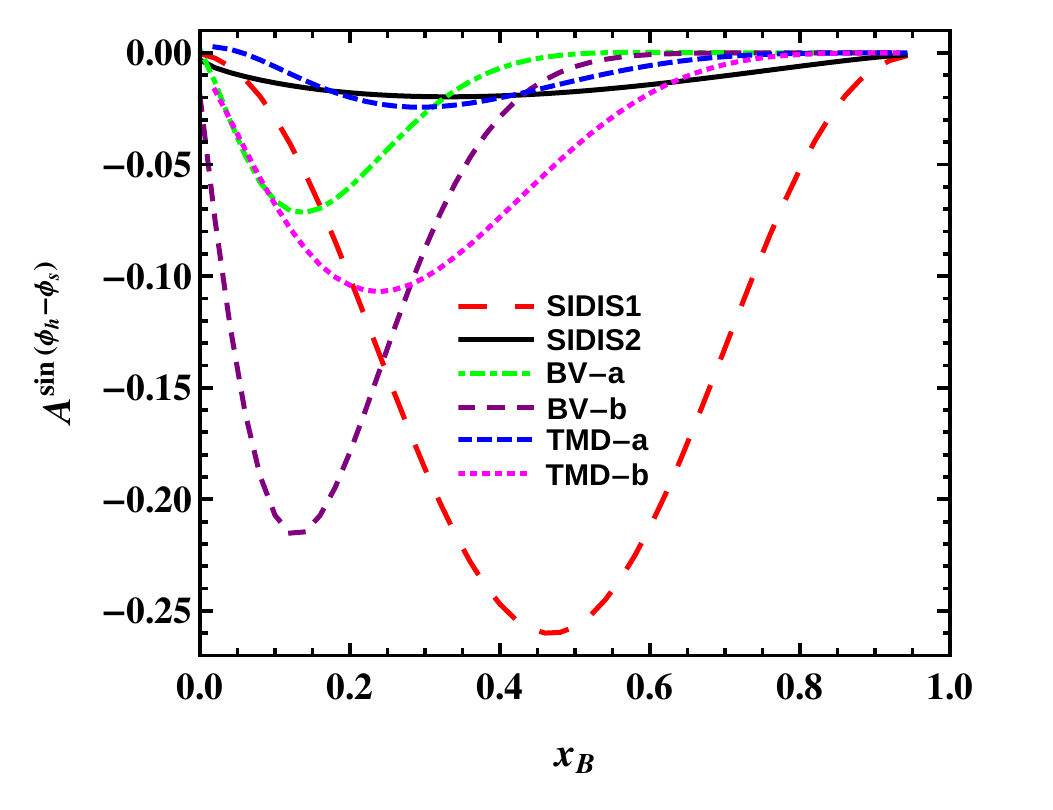}
\hspace{0.1cm}
\small{(b)}\includegraphics[width=8cm,height=6.5cm,clip]{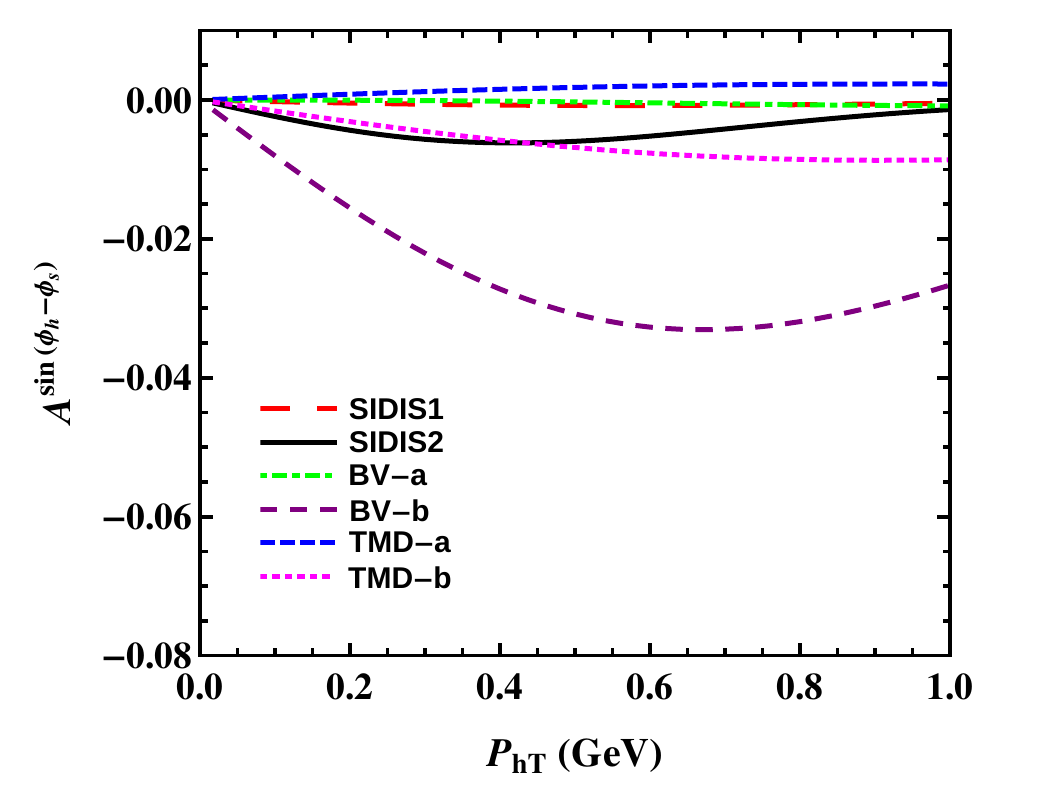}
\end{minipage}
\caption{\label{fig2}Single spin asymmetry  in $e+p^{\uparrow}\to e+J/\psi +X$
process as function of 
(a) $x_B$ (left panel) and  (b) $P_{hT}$ (right panel) at $\sqrt{s}=45.0$ GeV (EIC) using DGLAP (SIDIS1, SIDIS2,
BV-a and BV-b) and TMD (TMD-a and TMD-b) evolution approaches. The integration ranges are $0<P_{hT}<1.0$ GeV, 
$0.1<y<0.9$ and $0.0001<x_B<0.9$.}
\end{figure}
\begin{figure}[H]
\begin{minipage}[c]{0.99\textwidth}
\small{(a)}\includegraphics[width=8cm,height=6.5cm,clip]{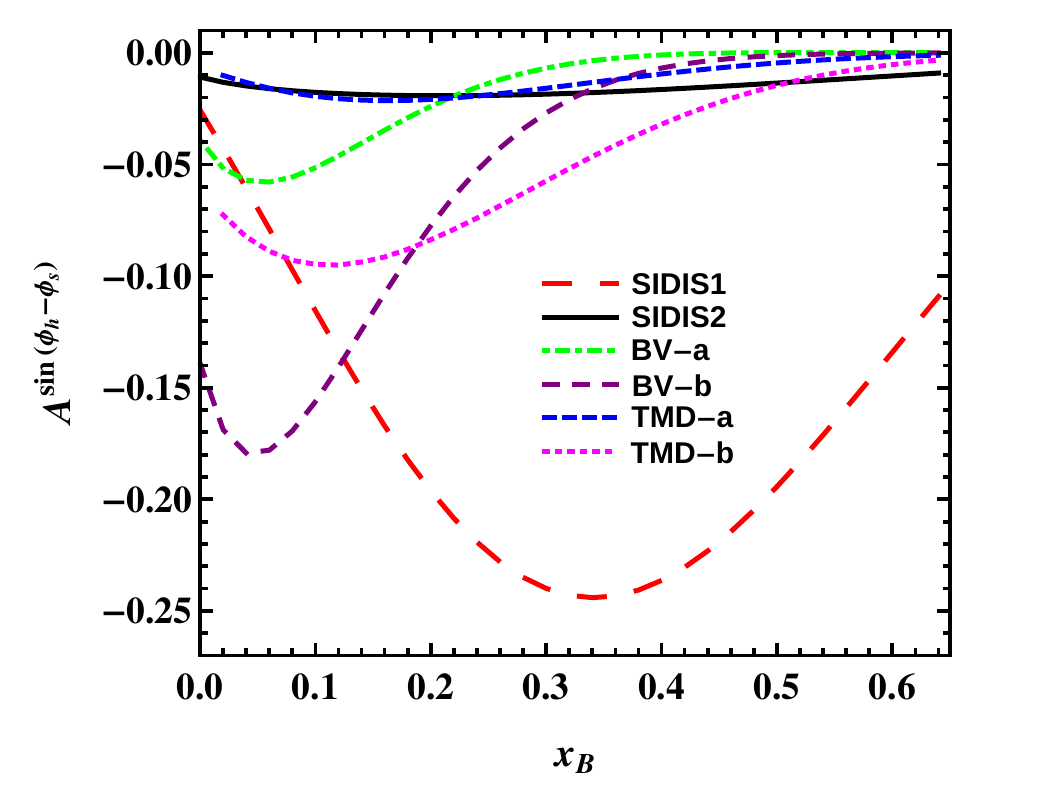}
\hspace{0.1cm}
\small{(b)}\includegraphics[width=8cm,height=6.5cm,clip]{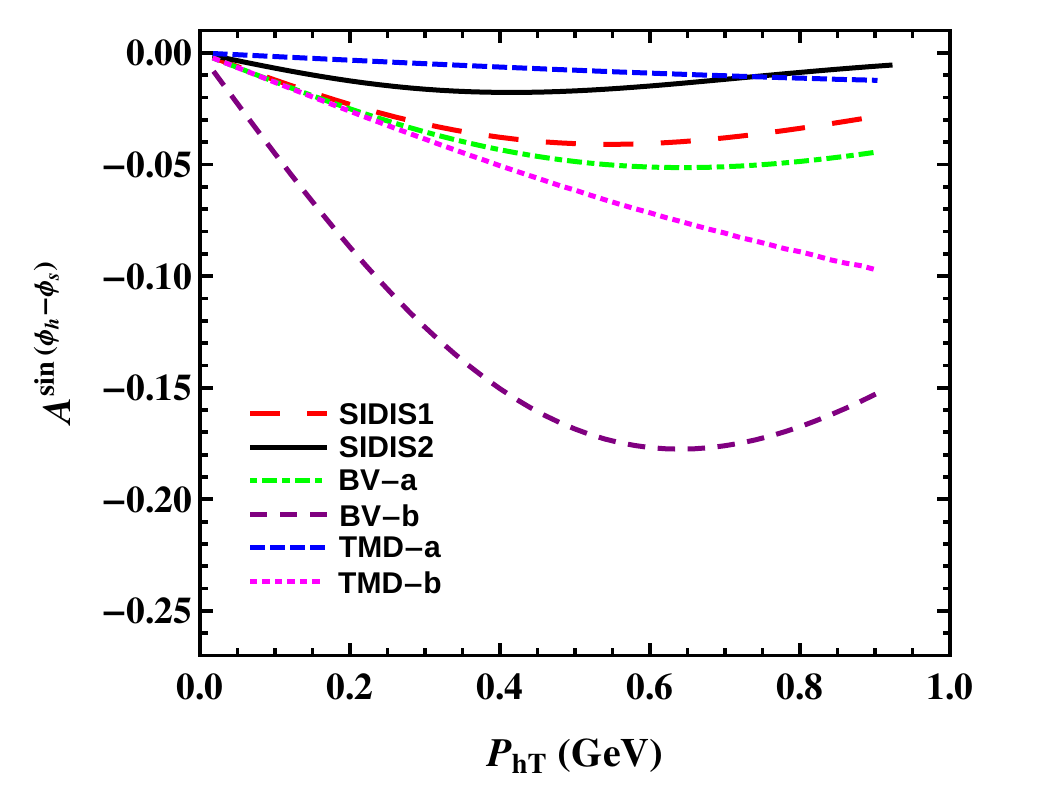}
\end{minipage}
\caption{\label{fig3}Single spin asymmetry  in $e+p^{\uparrow}\to e+J/\psi +X$
process as function of 
(a) $x_B$ (left panel) and  (b) $P_{hT}$ (right panel) at $\sqrt{s}=17.2$ GeV (COMPASS) using DGLAP (SIDIS1, SIDIS2,
BV-a and BV-b) and TMD (TMD-a and TMD-b) evolution approaches. The integration ranges are $0<P_{hT}<1.0$ GeV,
$0.1<y<0.9$ and $0.0001<x_B<0.65$.}
\end{figure}
\begin{figure}[H]
\begin{minipage}[c]{0.99\textwidth}
\small{(a)}\includegraphics[width=8cm,height=6.5cm,clip]{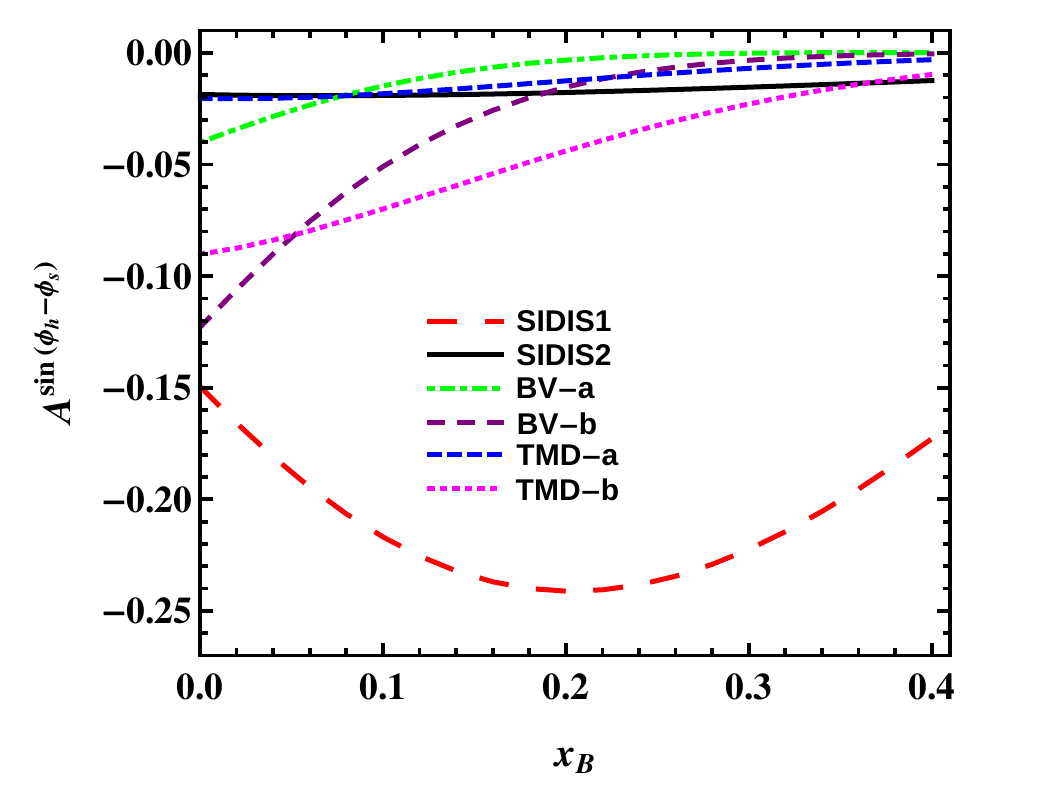}
\hspace{0.1cm}
\small{(b)}\includegraphics[width=8cm,height=6.5cm,clip]{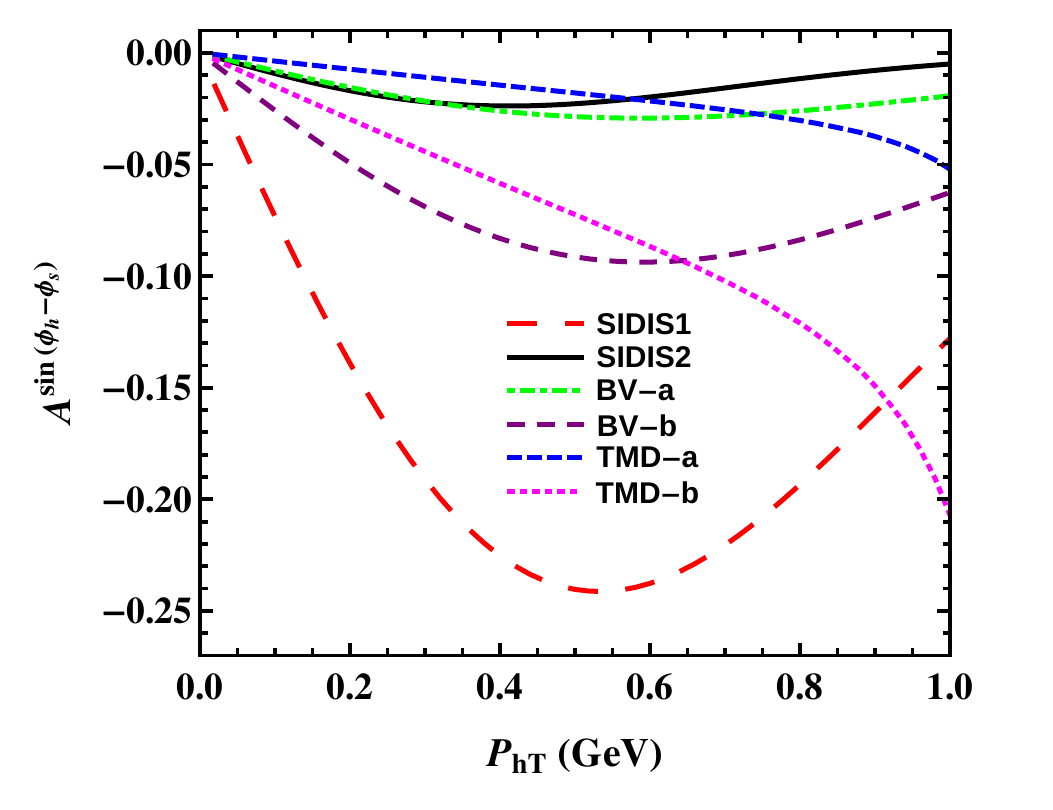}
\end{minipage}
\caption{\label{fig4}Single spin asymmetry  in $e+p^{\uparrow}\to e+J/\psi +X$
process as function of 
(a) $x_B$ (left panel) and  (b) $P_{hT}$ (right panel) at $\sqrt{s}=7.2$ GeV (HERMES) using DGLAP (SIDIS1, SIDIS2,
BV-a and BV-b) and TMD (TMD-a and TMD-b) evolution approaches. The integration ranges are $0<P_{hT}<1.0$ GeV, 
$0.35<y<0.95$ and
$0.023<x_B<0.40$.}
\end{figure}
\begin{figure}[H]
\begin{minipage}[c]{0.99\textwidth}
\small{(a)}\includegraphics[width=8cm,height=6.5cm,clip]{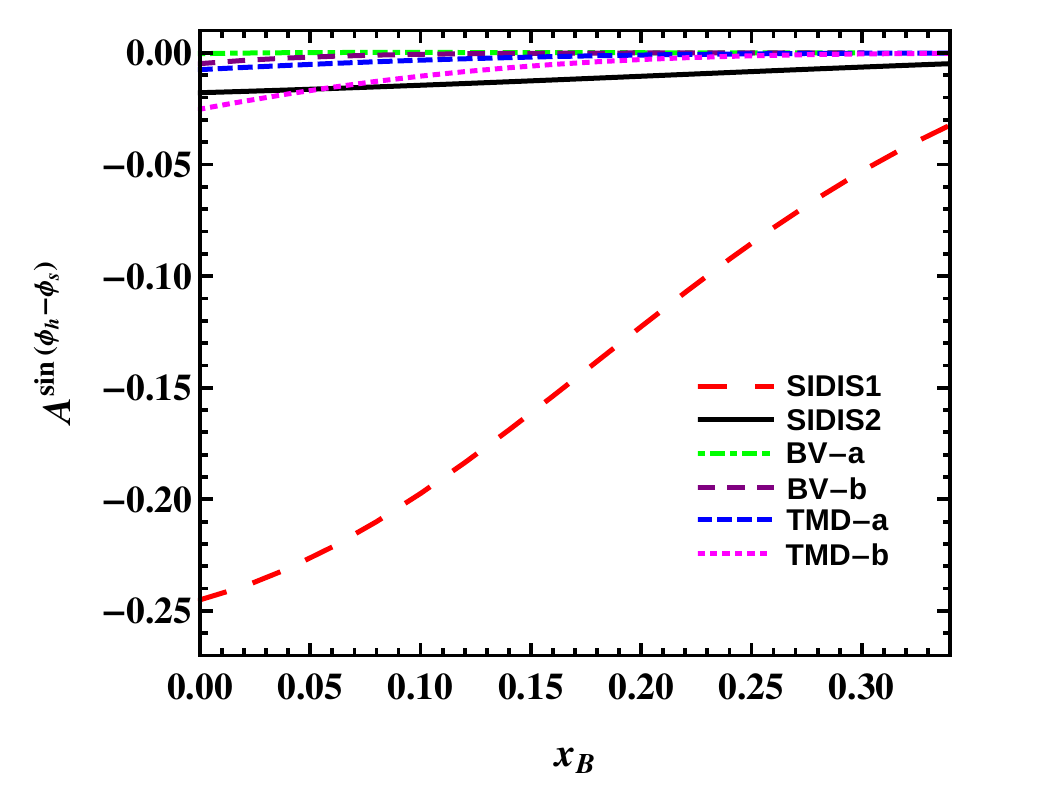}
\hspace{0.1cm}
\small{(b)}\includegraphics[width=8cm,height=6.5cm,clip]{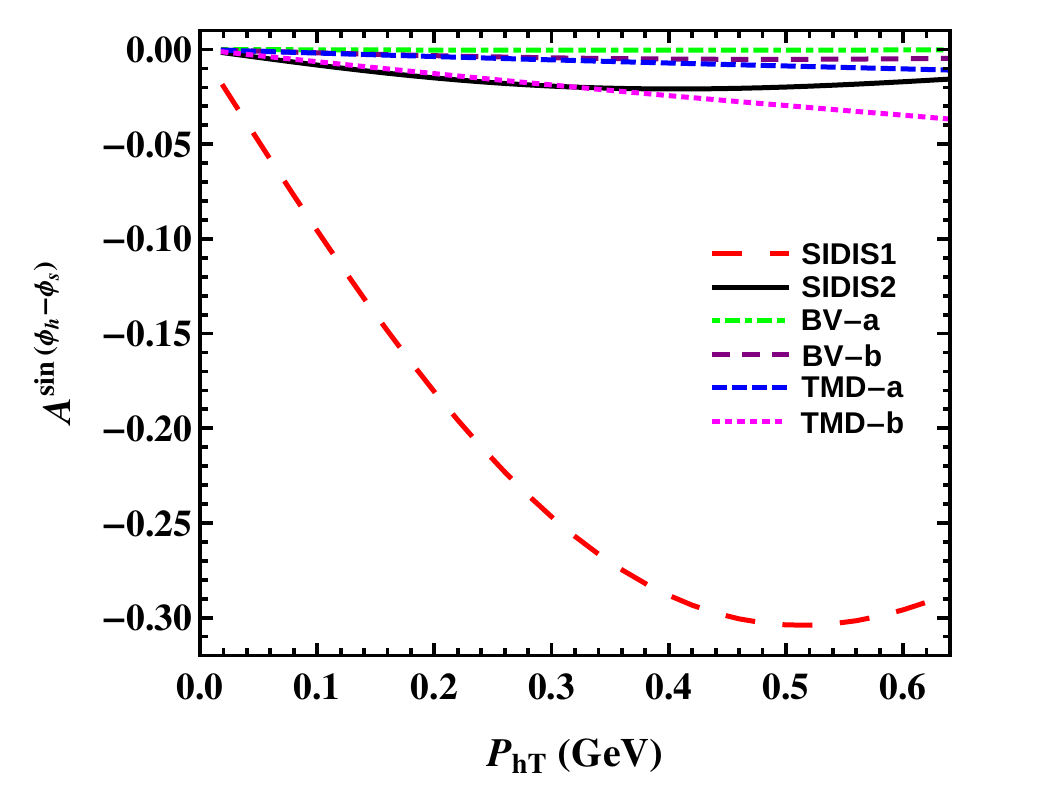}
\end{minipage}
\caption{\label{fig5}Single spin asymmetry  in $e+p^{\uparrow}\to e+J/\psi +X$
process as function of 
(a) $x_B$ (left panel) and  (b) $P_{hT}$ (right panel) at $\sqrt{s}=4.7$ GeV (JLab) using DGLAP (SIDIS1, SIDIS2,
BV-a and BV-b) and 
TMD (TMD-a and TMD-b) evolution approaches. The integration ranges are $0<P_{hT}<0.64$ GeV, $0.7<y<0.9$ 
and  
$0.0001<x_B<0.35$.}
\end{figure}

The following experimental cuts are imposed on the integration variables in Eq.\eqref{unp1}.
For COMPASS \cite{Adolph:2017pgv,Matousek:2016xbl}, 
$0.0001<x_B<0.65,~0.1<y<0.9~\mathrm{and}~0<P_{hT}<1.0~\mathrm{GeV}$, for HERMES \cite{Airapetian:2009ae},
$0.023<x_B<0.40,~0.35<y<0.95~\mathrm{and}~0<P_{hT}<1.0~\mathrm{GeV}$, for JLab \cite{Qian:2011py}
$0.0001<x_B<0.35,~0.7<y<0.9~\mathrm{and}~0<P_{hT}<0.64~\mathrm{GeV}$, and for EIC, 
$0.0001<x_B<0.9,~0.1<y<0.9~\mathrm{and}~0<P_{hT}<1.0~\mathrm{GeV}$.
The $\sin(\phi_h-\phi_s)$ weighted Sivers asymmetry for the kinematics of different experiments is shown 
in \figurename{\ref{fig2} 
-\ref{fig5}} as a function of $P_{hT}$ and $x_B$. The SSA is estimated both in DGLAP and 
Collins-Soper-Sterman (CSS) TMD evolution approach which is shown in \figurename{\ref{fig2}-\ref{fig5}}. 
The figures convention is as follows. \textquotedblleft SIDIS1\textquotedblright~and \textquotedblleft 
SIDIS2\textquotedblright~represent the SSA obtained in DGLAP evolution approach by considering two sets 
of best fit parameters SIDIS1 and SIDIS2 from Eq.\eqref{numer1} and \eqref{denom1}. Similarly,  
\textquotedblleft BV-a\textquotedblright~and \textquotedblleft BV-b\textquotedblright~represent the Sivers 
asymmetry obtained by using Eq.\eqref{ab} in DGLAP evolution. The obtained SSA in TMD evolution approach 
using two 
parameterizations from Eq.\eqref{ab} is denoted by \textquotedblleft TMD-a\textquotedblright~and 
\textquotedblleft TMD-b\textquotedblright.\par
Recently extracted gluon Sivers function \cite{alesio} from RHIC data and quark's Sivers function 
\cite{Anselmino:2016uie} from latest SIDIS data  have been employed in DGLAP evolution approach.
The SSA as a function of $P_{hT}$ is negative, and  is decreasing as the center of mass energy of the 
experiment increasing, which is maximum around $30\%$ at JLab energy. Moreover, Sivers asymmetry as a 
function of Bjorken variable ($x_B$) is negative and is maximum for SIDIS1 GSF parameters. Echevarria et 
al. \cite{echevarria}, have extracted $u$ and $d$ quark's
Sivers function by fitting data from JLab, HERMES and COMPASS within TMD evolution formalism. We use  
best fit parameters of these for gluon
Sivers function as defined in Eq.\eqref{ab} in CSS TMD evolution approach. Sivers asymmetry with 
respect to $P_{hT}$ obtained from SIDIS1 parameters is more at JLab and HERMES whereas SSA obtained from 
BV-b set parameters is dominant at COMPASS and EIC experiments. Basically, SSA is proportional  to gluon 
Sivers function which is considered as 
an average of $u$ and $d$ quark's $x$-dependent normalization $\mathcal{N}(x)$ in TMD-a 
parameterization. 
The sign of the asymmetry depends on relative magnitude of $N_u$ and $N_d$ and these have opposite sign which
can be observed in \tablename{ \ref{table1}}. Note that our kinematics is different from previous works in
\cite{Godbole:2012bx,Godbole:2013bca,Godbole:2014tha}, which also affects the sign. 
 The magnitude of $\mathcal{N}_u(x)$ is comparable but slightly dominant compared to 
$\mathcal{N}_d(x)$ at EIC $\sqrt{s}$. Therefore, the estimated Sivers asymmetry as a function of $P_{hT}$ 
using TMD-a parameters for EIC experiment is almost zero and positive. For JLab experiment, the 
estimated Sivers asymmetry by all the parameterizations except SIDIS1 is almost close to zero.\par
\begin{figure}[H]
\begin{center}
\includegraphics[width=8cm,height=6.5cm,clip]{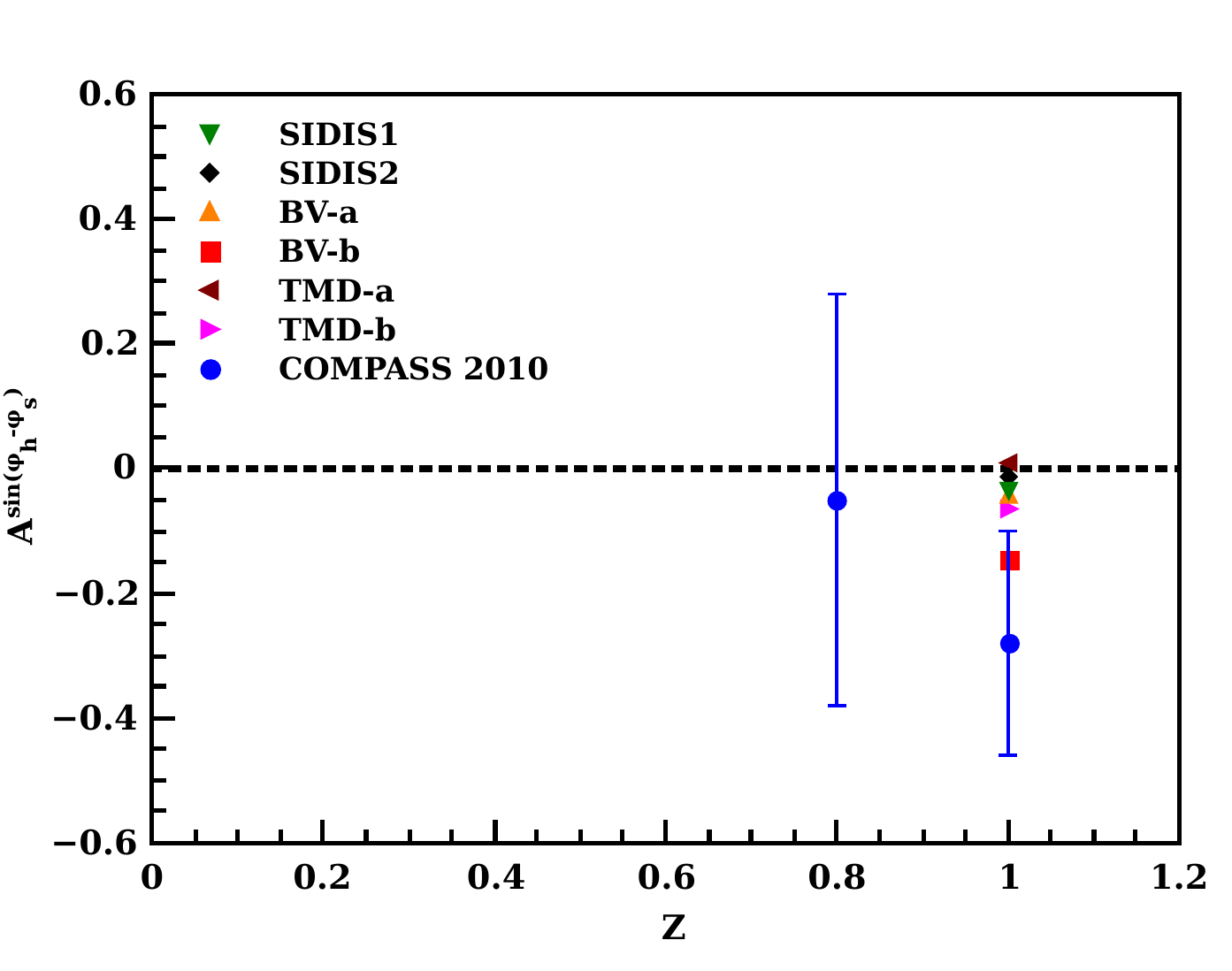}
\end{center}

\caption{\label{fig6}Single spin asymmetry  in $e+p^{\uparrow}\to e+J/\psi +X$
process at $z=1$ with $\sqrt{s}=17.2$ GeV (COMPASS) using DGLAP (SIDIS1, SIDIS2, BV-a and BV-b) and 
TMD (TMD-a and TMD-b) evolution approaches. The integration ranges are $0<P_{hT}<1.0$ GeV, $0.1<y<0.9$ and 
$0.0001<x_B<0.65$. Data from \cite{Matousek:2016xbl}.}
\end{figure}

The delta function in Eq.\eqref{delta} implies that $z=1$ (LO). In \figurename{~\ref{fig6}}, the obtained 
Sivers asymmetry at $z=1$ is compared with COMPASS data \cite{Matousek:2016xbl}. Interestingly, all the 
set of parameters  give negative asymmetry. However, estimated SSA  with BV-b set of parameters is 
within the error bar of the experiment. In Ref. \cite{Adolph:2017pgv}, negative gluon Sivers asymmetry with more than two 
standard deviation, $A^{Siv}_{PGF}=-0.23\pm0.08$, is reported in SIDIS process based on Monte carlo 
simulation analysis.   As stated before, it is expected that the Sivers 
function has different sign in DY and SIDIS process, which comes from the gauge link.  Sivers function in SIDIS has 
been extracted by COMPASS \cite{Adolph:2017pgv,Matousek:2016xbl},  HERMES \cite{Airapetian:2009ae} and 
JLab \cite{Qian:2011py} collaboration. However, information about the DY Sivers function has not been 
explored, since polarized DY process has not been measured ever. Only very recently, data is available in DY process
$pp^\uparrow\rightarrow W^{\pm}/Z+X$ \cite{Adamczyk:2015gyk}. Anselmino et al. \cite{Anselmino:2016uie} 
have first time attempted to study the nonuniversality signature i.e., sign change of Sivers function, 
however, they could not draw a definite conclusion about it due to poor data, although data for
 $W^-$ production seem to favor the sign change. \par
\begin{figure}[H]
\begin{minipage}[c]{0.99\textwidth}
\small{(a)}\includegraphics[width=8cm,height=6.5cm,clip]{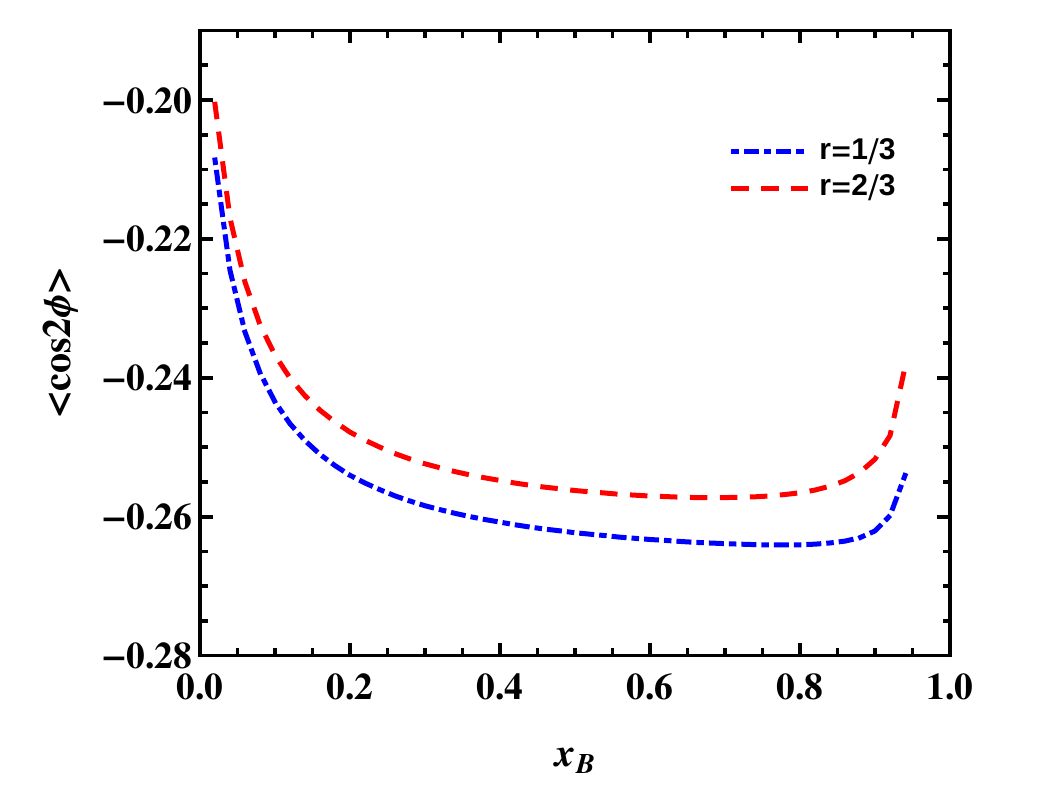}
\hspace{0.1cm}
\small{(b)}\includegraphics[width=8cm,height=6.5cm,clip]{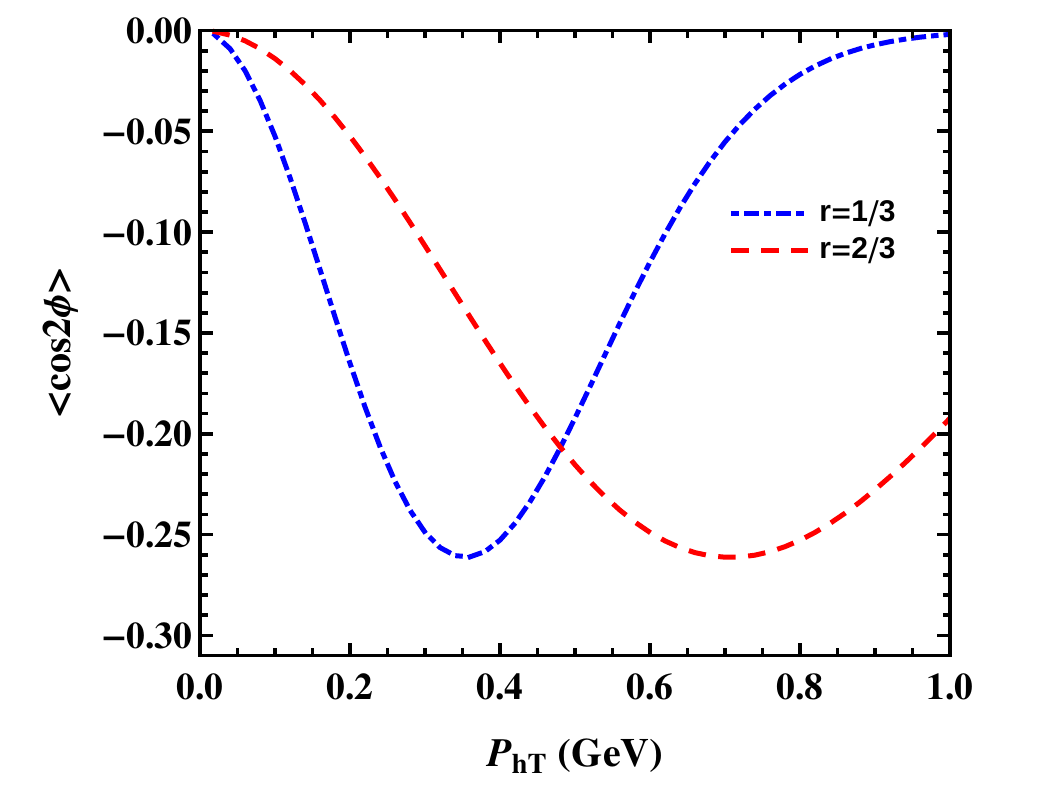}
\end{minipage}
\caption{\label{fig7}$\cos2\phi$ asymmetry  in $e+p\rightarrow e+J/\psi +X$
process as function of 
(a) $x_B$ (left panel) and  (b) $P_{hT}$ (right panel) at $\sqrt{s}=45.0$ GeV (EIC). The integration ranges
are $0<P_{hT}<1.0$ GeV, $0.1<y<0.9$ and $0.0001<x_B<0.9$.}
\end{figure}
\begin{figure}[H]
\begin{minipage}[c]{0.99\textwidth}
\small{(a)}\includegraphics[width=8cm,height=6.5cm,clip]{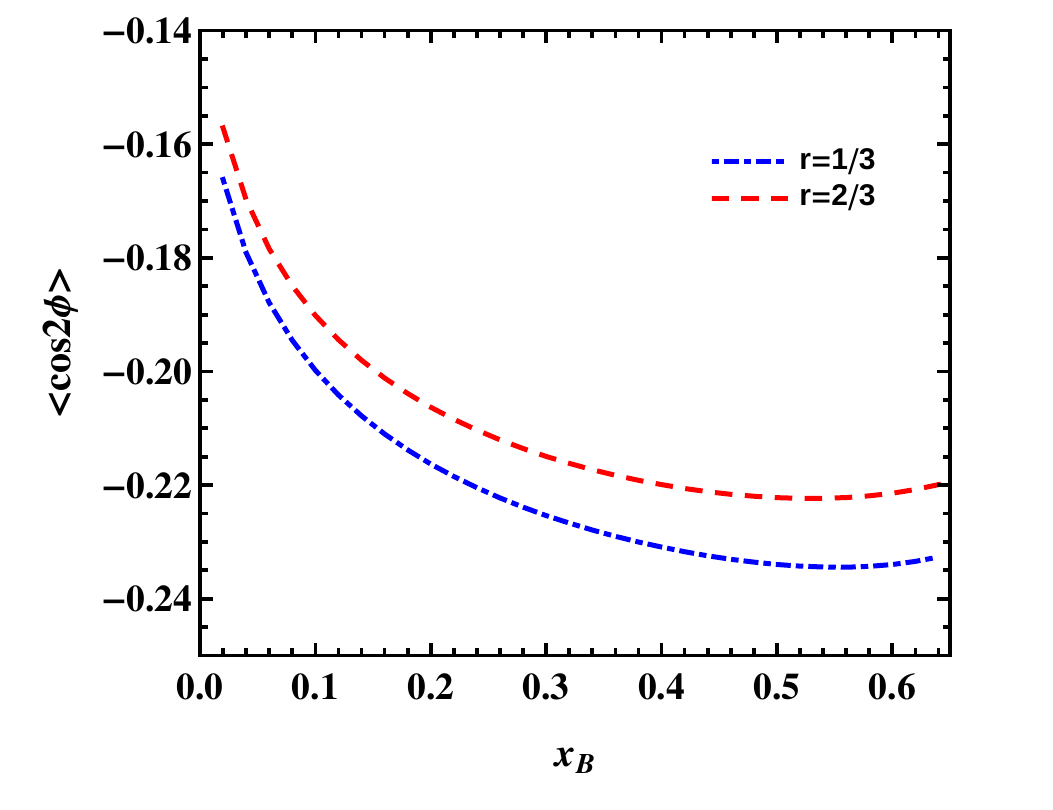}
\hspace{0.1cm}
\small{(b)}\includegraphics[width=8cm,height=6.5cm,clip]{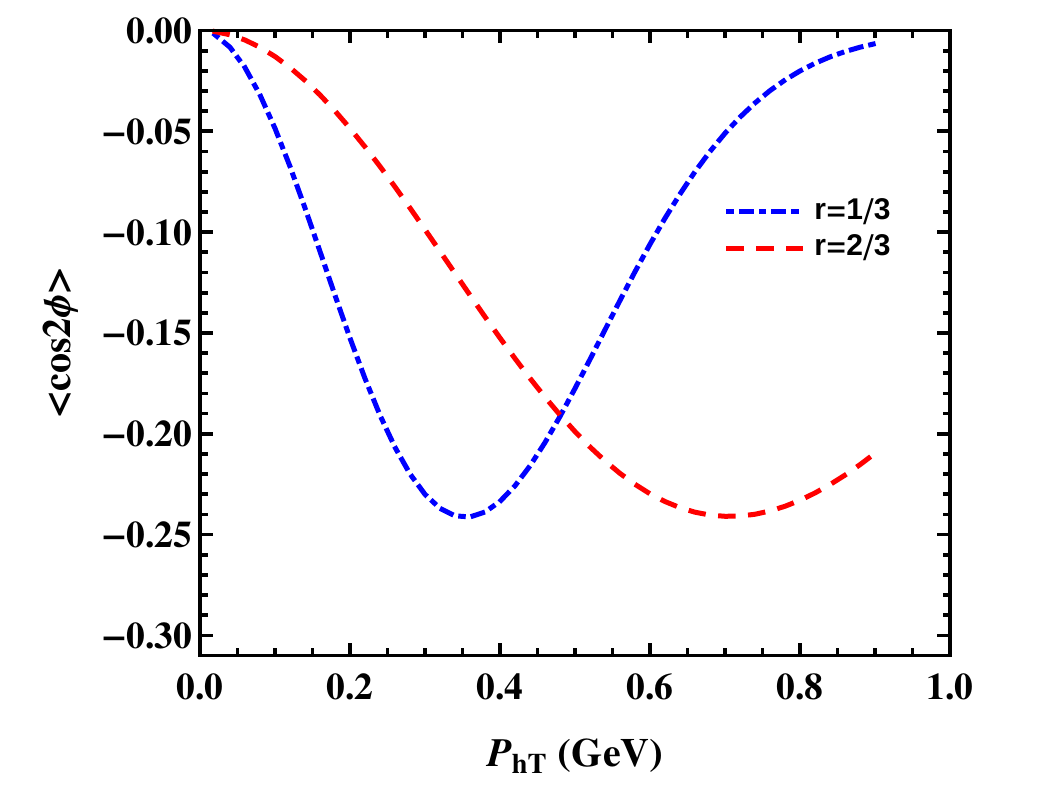}
\end{minipage}
\caption{\label{fig8}$\cos2\phi$ asymmetry  in $e+p\rightarrow e+J/\psi +X$
process as function of (a) $x_B$ (left panel) and  (b) $P_{hT}$ (right panel) at $\sqrt{s}=17.2$ GeV 
(COMPASS). The integration ranges are $0<P_{hT}<1.0$ GeV, $0.1<y<0.9$ and $0.0001<x_B<0.65$.}
\end{figure}
\begin{figure}[H]
\begin{minipage}[c]{0.99\textwidth}
\small{(a)}\includegraphics[width=8cm,height=6.5cm,clip]{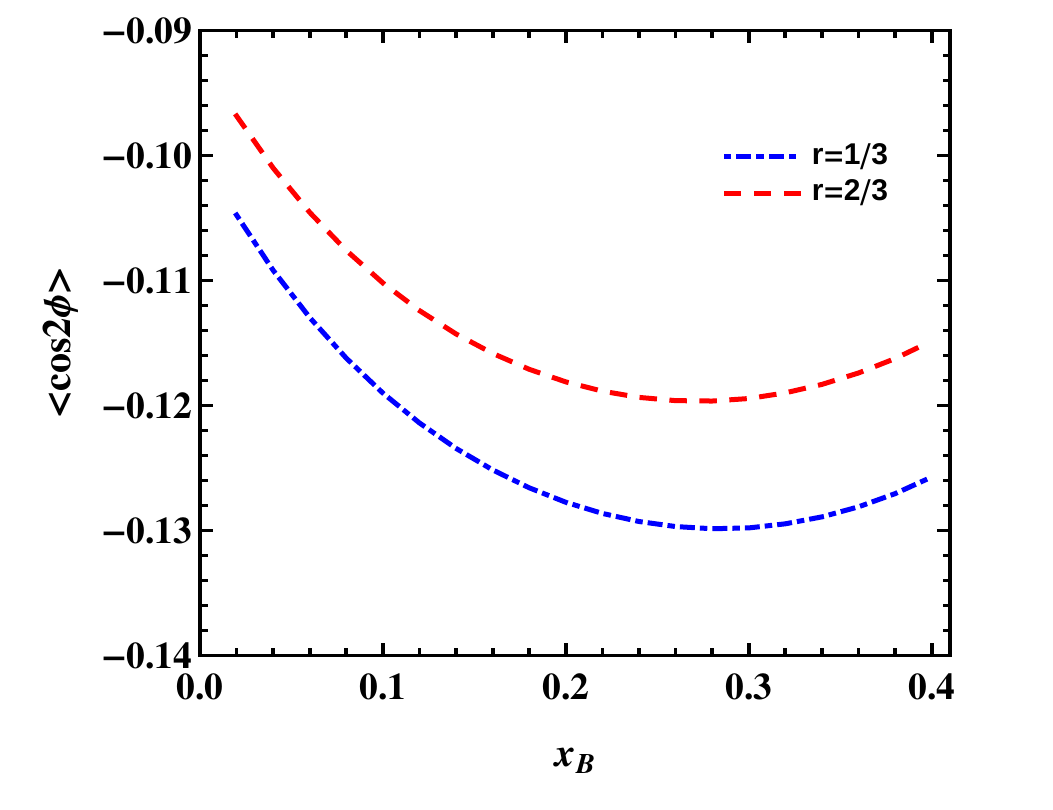}
\hspace{0.1cm}
\small{(b)}\includegraphics[width=8cm,height=6.5cm,clip]{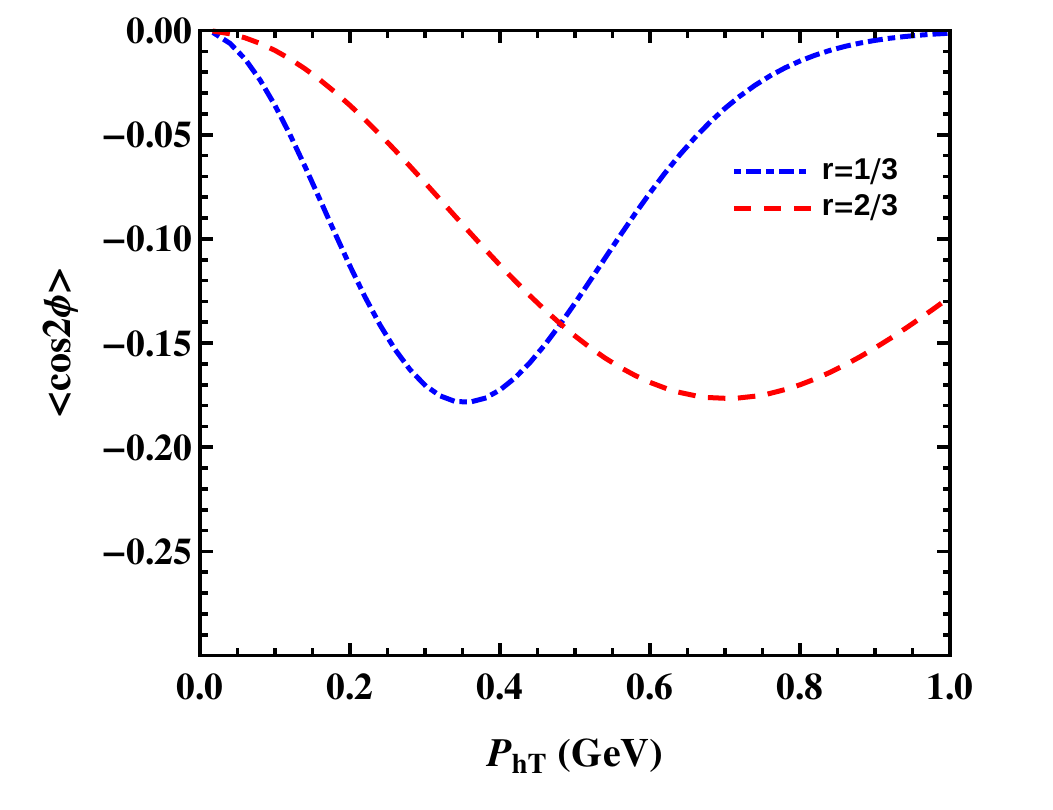}
\end{minipage}
\caption{\label{fig9}$\cos2\phi$ asymmetry  in $e+p\rightarrow e+J/\psi +X$
process as function of (a) $x_B$ (left panel) and  (b) $P_{hT}$ (right panel) at $\sqrt{s}=7.2$ GeV 
(HERMES). The integration ranges are $0<P_{hT}<1.0$ GeV, $0.35<y<0.95$ and $0.023<x_B<0.40$.}
\end{figure}
\begin{figure}[H]
\begin{minipage}[c]{0.99\textwidth}
\small{(a)}\includegraphics[width=8cm,height=6.5cm,clip]{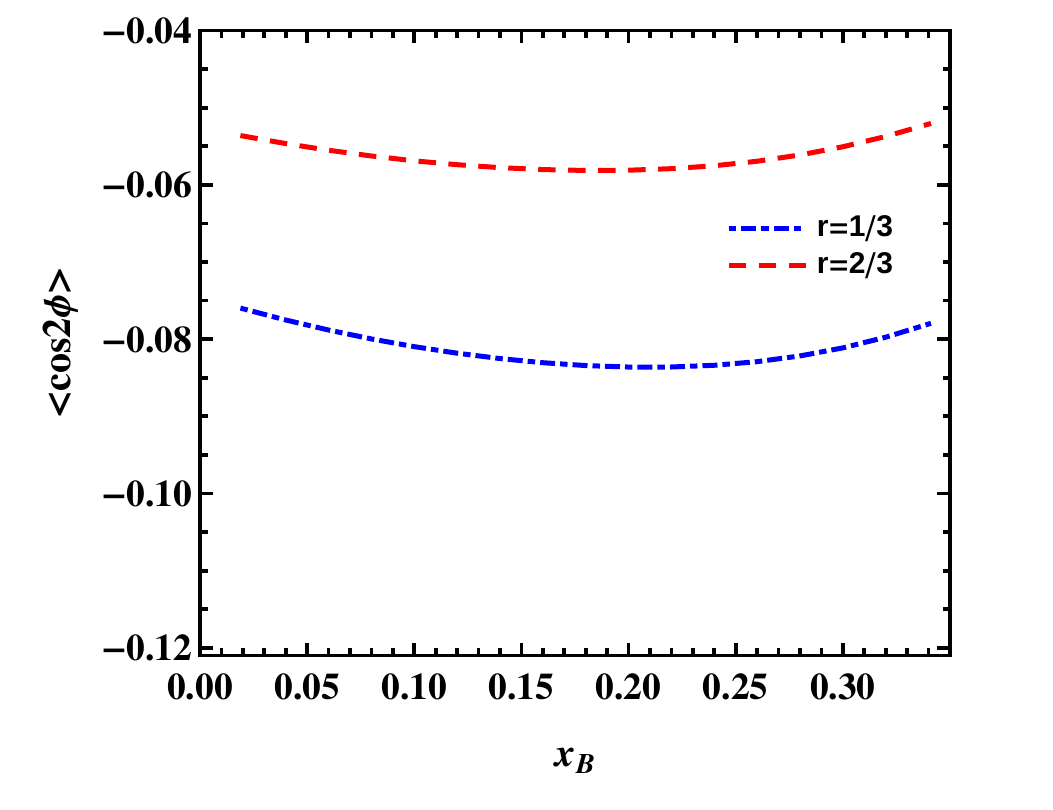}
\hspace{0.1cm}
\small{(b)}\includegraphics[width=8cm,height=6.5cm,clip]{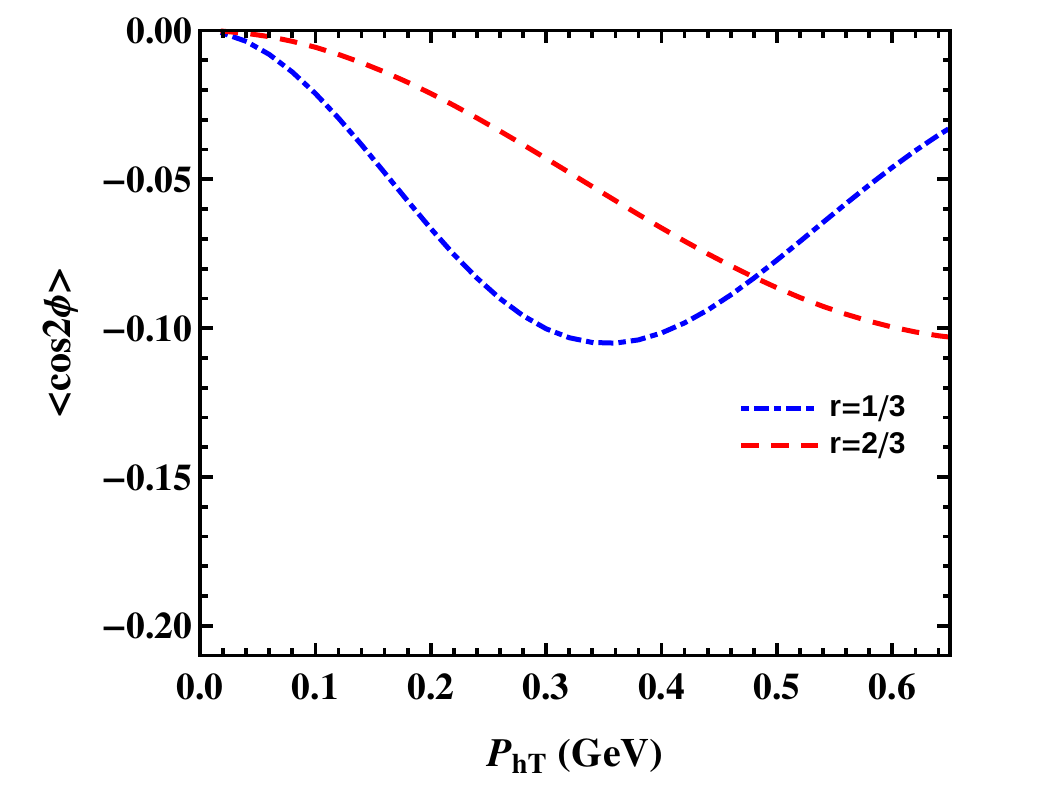}
\end{minipage}
\caption{\label{fig10}$\cos2\phi$ asymmetry  in $e+p\rightarrow e+J/\psi +X$
process as function of (a) $x_B$ (left panel) and  (b) $P_{hT}$ (right panel) at $\sqrt{s}=4.7$ GeV 
(JLab). The integration ranges are $0<P_{hT}<0.64$ GeV, $0.7<y<0.9$ and $0.0001<x_B<0.35$.}
\end{figure}

The $\cos2\phi$ asymmetry is shown in \figurename{ \ref{fig7}-\ref{fig10}} as a function of $x_B$ 
and $P_{hT}$ for $r=1/3$ and $r=2/3$. To obtain $\cos2\phi$ asymmetry, the Gaussian parameterizations 
for unpolarized and linearly polarized gluon distribution functions are used, as defined in 
Eq.\eqref{unp} and \eqref{hg}. Until now, experimental investigation has not been done to extract the 
unknown Boer-Mulders function, $h^{\perp g}_1$. In Ref.  \cite{Mukherjee:2015smo,Mukherjee:2016cjw}, the 
effect of $h^{\perp g}_1$ on the unpolarized differential cross section of $J/\psi$ production in $pp$ 
collision is explored. The $J/\psi$ production in unpolarized $ep$ collision process is also a reliable channel to probe the 
$h^{\perp g}_1$ by measuring $\cos2\phi$ asymmetry. 
It is obvious from Eq.\eqref{bb0} that the negative $\cos2\phi$ asymmetry as function of $x_B$ and $P_{hT}$ 
is obtained due to the dominant contribution of $\leftidx{^{1}}{S}{_0}$ state compared to the other states 
($\leftidx{^{3}}{P}{_0}$, $\leftidx{^{3}}{P}{_1}$ and $\leftidx{^{3}}{P}{_2}$).
$\cos2\phi$ asymmetry as a function of $P_{hT}$ is almost same 
for all the experiments, however, maximum value of $<\cos2\phi>$  decreases with $\sqrt{s}$.
The maximum of $26\%$ $\cos2\phi$ asymmetry as a function of $x_B$ is observed at EIC experiment.


\section{Conclusion}\label{sec6}
We have calculated the Sivers  and $\cos2\phi$ asymmetries in the  production of $J/\psi$ in 
polarized and unpolarized $ep$ collision respectively. $J/\psi$ production
process gives direct access to the gluon Sivers function at leading order through the channel 
$\gamma^\ast g \rightarrow c \bar{c}$. We used the NRQCD based color octet model and a formalism based on
TMD factorization. Sizable negative Sivers asymmetry is observed in $J/\psi$ production. The estimated 
SSA at $z=1$ is compared 
with COMPASS data and is in considerable agreement. We investigated the effect of TMD evolution on the 
 Sivers asymmetry. Moreover, Sizable $\cos2\phi$ asymmetry is obtained in unpolarized SIDIS process 
which allows to probe the Boer-Mulders function, $h^{\perp g}_1$.   Thus the asymmetries in the polarized 
and unpolarized SIDIS  processes  are  important observables to give valuable  information on the gluon 
 Sivers function and linearly polarized gluon TMD respectively. Further work would involve taking into 
account 
higher order corrections to the asymmetry, where effect of the charmonium production mechanism is likely to 
play an important  role. 
\section*{Acknowledgment}
We would like to thank Mauro Anselmino for fruitful discussion during his stay at IIT Bombay.  Cristian 
Pisano is thanked for useful discussion.
\appendix*
 \begin{center}
 \section*{Appendix: {LO\lowercase{ amplitude of}} $\gamma^{\ast}g\rightarrow J/\psi$}
\end{center}
\begin{figure}[H]
\begin{center} 
\includegraphics[height=3.5cm,width=12cm]{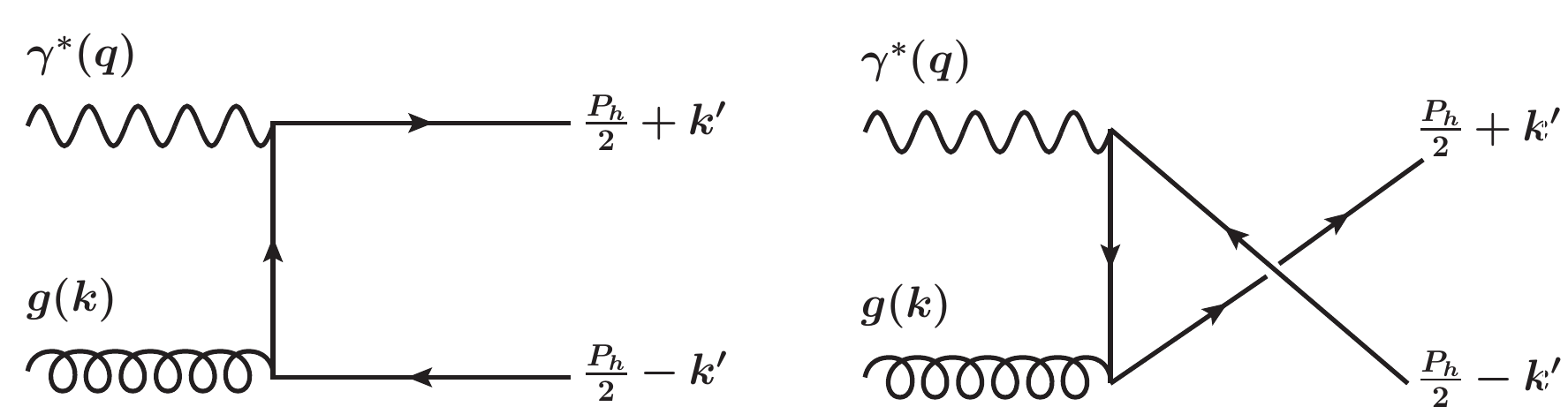}
\end{center}
\caption{\label{fig11}Feynman diagrams for  $\gamma^{\ast}+g\rightarrow J/\psi$ process.}
\end{figure}
As per Ref. \cite{Baier:1983va,Boer:2012bt}, the amplitude of the quarkonium bound state can be written as 
bellow
\begin{equation}\label{a1}
 \begin{aligned}
 \mathcal{M}^{\mu\nu}\left(\gamma^{\ast}g \rightarrow  
Q\bar{Q}[\leftidx{^{2S+1}}{L}{_J}^{(1,8a)}]\right)={}&\sum_{L_zS_z}
 \int \frac{d^3\bm{k}^{\prime}}{(2\pi)^3}\Psi_{LL_z}(\bm{k}^\prime)
 \langle LL_z;SS_z|JJ_z\rangle \mathrm{Tr}[O^{\mu\nu}(q,k,P_h,k^\prime)\\
 &\times\mathcal{P}_{SS_z}(P_h,k^\prime)]
 \end{aligned}
\end{equation}
where $k^\prime$ is the relative momentum of the heavy quark in the quarkonium rest frame. 
The eigenfunction of the orbital angular momentum $L$ is $\Psi_{LL_z}(\bm{k}^\prime)$. We follow the similar 
calculation as reported in \cite{Boer:2012bt}, hence only the important steps are presented below and for 
more details Ref.\cite{Boer:2012bt} is preferred. From 
\figurename{\ref{fig11}}, the amplitude of heavy quark pair is given by
\begin{equation}\label{a2}
 \begin{aligned}
O^{\mu\nu}(q,k,P_h,k^\prime)=&{}\sum_{ij}\langle 3i;\bar{3}j|8a\rangle g_s(ee_c)
\Bigg\{\gamma^\nu\frac{\slashed{P_h}/2+\slashed{k}^\prime-\slashed{q}+m_c}{(P_h/2+k^\prime-q)^2-m_c^2}
\gamma^\mu(T^b)^{ji}\\
&+\gamma^\mu(T^b)^{ji}\frac{\slashed{P_h}/2+\slashed{k}^\prime-\slashed{k}+m_c}{(P_h/2+k^\prime-k)^2-m_c^2}
\gamma^\nu\Bigg\}
\end{aligned}
\end{equation}
The sum over the SU(3) Clebsch-Gordan coefficients project out  the color state of $Q\bar{Q}$ pair either it 
is in color singlet or octet state, and  are defined as 
$\langle 3i;\bar{3}j|1\rangle=\frac{\delta^{ij}}{\sqrt{N_c}}~,
\langle 3i;\bar{3}j|8a\rangle=\sqrt{2}(T^a)^{ij}$ for color singlet and color octet states respectively. 
$T^b$ is the SU(3) Gell-Mann matrix. Charm quark and quarkonium bound state masses are denoted with $m_c$ 
and $M=2m_c$ respectively.
The excluded external legs in Eq.\eqref{a2} are absorbed in the spin projection operator which is given by 
\be \label{a3}
\mathcal{P}_{SS_z}(P_h,k^\prime)&=&\sum_{s_1s_2}\langle\frac12s_1;\frac12s_2|SS_z\rangle 
\upsilon(\frac{P_h}{2}-
k^\prime,s_2)\bar{u}(\frac{P_h}{2}+k^\prime,s_1)\nonumber\\
&=&\frac{1}{4M^{3/2}}(-\slashed{P}_h+2\slashed{k}^\prime+M)\Pi_{SS_z}(\slashed{P}_h+2\slashed{k}^\prime+M)
+\mathcal{O}(k^{\prime 2})
\ee
bear with $\Pi_{SS_z}=\gamma^{5}$ for singlet ($S=0$) state and $\Pi_{SS_z}=\slashed{\varepsilon}_{s_z}(P_h)$
for triplet ($S=1$) state. Here spin polarization vector of the $Q\bar{Q}$ system is denoted with
$\varepsilon_{s_z}(P_h)$. The Taylor expansion around $\bm{k}^\prime=0$ in Eq.\eqref{a1} gives the $S$-wave 
and $P$-wave amplitudes. The first term in the expansion is $S$-wave amplitude
\begin{eqnarray}\label{a4}
 \mathcal{M}^{\mu\nu}[\leftidx{^{1}}{S}{_0}^{(8a)}]=\frac{1}{4\sqrt{\pi 
M}}R_0(0)\mathrm{Tr}[O^{\mu\nu}(0)(-\slashed{P}_h+M)\gamma^5]
\end{eqnarray}
and
\begin{equation}\label{a5}
 \mathcal{M}^{\mu\nu}[\leftidx{^{3}}{S}{_1}^{(8a)}]=\frac{1}{4\sqrt{\pi 
M}}R_0(0)\mathrm{Tr}[O^{\mu\nu}(0)(-\slashed{P}_h+M)\slashed{\varepsilon}_{s_z}].
\end{equation}
 The derivative term in the expansion of Eq.\eqref{a1} is the $P$-wave amplitude
 \begin{eqnarray}\label{a6} 
  \begin{aligned}
 \mathcal{M}^{\mu\nu}[\leftidx{^{3}}{P}{_0}^{(8a)}]={}&-\frac{i}{\sqrt{4\pi M}}R^\prime_1(0)
 \mathrm{Tr}\Big[3O^{\mu\nu}(0)+\left(\gamma_\alpha 
O^{\mu\nu\alpha}(0)+\frac{P_{h\alpha}}{M}O^{\mu\nu\alpha}(0)\right)\\
&\times\frac{-\slashed{P}_h+M}{2} \Big],
 \end{aligned}
 \end{eqnarray}
 \begin{eqnarray}\label{a7} 
 \begin{aligned}
 \mathcal{M}^{\mu\nu}[\leftidx{^{3}}{P}{_1}^{(8a)}]={}&-\sqrt{\frac{3}{8\pi M}}R^\prime_1(0)
\epsilon_{\rho\sigma\alpha\beta}\frac{P_h^\rho}{M}
 \varepsilon^\sigma_{J_z}(P_h)
\mathrm{Tr}\Big[\gamma^\alpha O^{\mu\nu\beta}(0)\frac{-\slashed{P}_h+M}{2}\\&-O^{\mu\nu}(0) \frac{\slashed{P}_h}{M}\gamma^\alpha\gamma^\beta\Big]
 \end{aligned}
 \end{eqnarray}
 and 
 \begin{eqnarray}\label{a8} 
 \mathcal{M}^{\mu\nu}[\leftidx{^{3}}{P}{_2}^{(8a)}]=-i\sqrt{\frac{3}{4\pi M}}R^\prime_1(0)
 \varepsilon^{\alpha\beta}_{ J_z }(P_h)
 \mathrm{Tr}\left[\gamma_\beta O^{\mu\nu}_\alpha(0)\frac{-\slashed{P}_h+M}{2}\right]
\end{eqnarray}
The definitions of $O^{\mu\nu}(0)$ and $O^{\mu\nu\alpha}(0)$ are obtained from Eq.\eqref{a2} which are given by
\begin{equation}\label{a9}
 \begin{aligned}
O^{\mu\nu}(0)= \frac{\sqrt{2}g_s(ee_c)\delta^{ab}}{2(q^2-M^2)}
\Big\{\gamma^\nu\left(\slashed{P_h}-2\slashed{q}+M\right)\gamma^\mu
+\gamma^\mu\left(\slashed{P_h}-2\slashed{k}+M\right)\gamma^\nu\Big\},
\end{aligned}
\end{equation}
\begin{equation}\label{a10}
\begin{aligned}
O^{\mu\nu\alpha}(0)=\frac{\partial}{\partial k^{\prime}_{ 
\alpha}}O(q,k,P_h,k^\prime)\Big\rvert_{k^\prime=0}=&{}\frac{\sqrt{2}g_s(ee_c)\delta^{ab}}{
(q^2-M^2)}\Bigg\{\
\frac{2k^\alpha}{q^2-M^2}\Big[\gamma^\mu\left(\slashed{P_h}-2\slashed{k}
+M\right)\gamma^\nu\\
&+\gamma^\nu\left(\slashed{P_h}-2\slashed{k}-M\right)\gamma^\mu\Big]
+\gamma^\mu\gamma^\alpha\gamma^\nu+\gamma^\nu\gamma^\alpha\gamma^\mu
\Bigg\}.
\end{aligned}
\end{equation}
Here $R_0(0)$ and $R_1^\prime(0)$ are the radial wave function and its derivative at the origin, and have the 
following relation with LDME \cite{Ko:1996xw}
\be\label{a11}
\langle 0\mid \mathcal{O}_8^{J/\psi}(\leftidx{^1}{S}{_J})\mid 0\rangle=\frac{2}{\pi}(2J+1)|R_0(0)|^2
\ee
\be\label{a12}
\langle 0\mid \mathcal{O}_8^{J/\psi}(\leftidx{^3}{P}{_J})\mid 0\rangle=\frac{2N_c}{\pi}(2J+1)|R^\prime_1(0)|^2
\ee
 After taking the trace one obtains the following amplitude expressions for $S$-wave and $P$-wave states
 \be\label{a13}
\mathcal{M}^{\mu\nu}[\leftidx{^{1}}{S}{_0}^{(8a)}]=2i\frac{\sqrt{2}g_s(ee_c)\delta^{ab}}{\sqrt{\pi 
M}(Q^2+M^2)}R_0(0)
\epsilon^{\mu\nu\rho\sigma}k_{\rho}P_{h\sigma}
\ee
\be\label{a14}
\mathcal{M}^{\mu\nu}[\leftidx{^{3}}{S}{_1}^{(8a)}]=\frac{\sqrt{2}g_s(ee_c)\delta^{ab}}{\sqrt{\pi 
M}(Q^2+M^2)}R_0(0)
4Mg^{\mu\nu}P_h^\beta 
\varepsilon_{s_z\beta}(P_h)=0
\ee
\begin{eqnarray}\label{a15} 
\begin{aligned}
 \mathcal{M}^{\mu\nu}[\leftidx{^{3}}{P}{_0}^{(8a)}]=&{}2i\frac{\sqrt{2}g_s(ee_c)\delta^{ab}}{\sqrt{\pi 
M^3}}
 R^\prime_1(0)\frac{3M^2+Q^2}{M^2+Q^2}\Big[
g^{\mu\nu}-2\frac{k^\nu P^\mu_h}{M^2+Q^2}\Big]
\end{aligned}
\end{eqnarray}
\begin{eqnarray}\label{a16} 
\begin{aligned}
 \mathcal{M}^{\mu\nu}[\leftidx{^{3}}{P}{_1}^{(8a)}]={}&\sqrt{\frac{3}{8\pi 
M}}\frac{\sqrt{2}g_s(ee_c)\delta^{ab}}
 {Q^2+M^2} R^\prime_1(0)
 \epsilon_{\rho\sigma\alpha\beta}\frac{P_h^\rho}{M}
 \varepsilon^\sigma_{J_z}(P_h) 
\frac{4}{M}\Bigg\{g^{\mu\beta}\Big((M^2-Q^2)\\
&\times g^{\nu\alpha}+2k^{\alpha}P_h^\nu\Big)
+g^{\nu\beta}\left((M^2+Q^2)g^{\mu\alpha}-2k^{\alpha}P_h^\mu\right)-2g^{\mu\alpha}k^\beta 
P_h^{\nu}\\
&-\frac{2k^\beta}{M^2+Q^2}
\left(2M^2g^{\mu\nu} k^\alpha -2M^2g^{\mu\alpha} 
k^\nu+(M^2-Q^2)g^{\nu\alpha}P_h^\mu\right)
\Bigg\}
\end{aligned}
 \end{eqnarray}
\begin{eqnarray}\label{a17}
\begin{aligned}
 \mathcal{M}^{\mu\nu}[\leftidx{^{3}}{P}{_2}^{(8a)}]=&{}2i\sqrt{\frac{3}{\pi M}}\frac{\sqrt{2}g_s(ee_c)\delta^{ab}M}
 {(Q^2+M^2)} R^\prime_1(0)\varepsilon_{J_z\alpha\beta}(P_h)
\Big[g^{\alpha\nu}g^{\beta\mu}+g^{\alpha\mu}g^{\beta\nu}\\
&-\frac{4k^\alpha }{Q^2+M^2}\left(k^\beta g^{\mu\nu}
-k^\nu g^{\beta\mu}+P_h^\mu g^{\beta\nu}\right) \Big]
\end{aligned}
\end{eqnarray}

\bibliographystyle{apsrev}
\bibliography{rf}

\end{document}